\documentclass[twocolumn,twoside,slac_two]{revtex4}
\usepackage{graphicx}
\usepackage{fancyhdr}
\begin{document}

\title{Cosmic Gamma-ray Background Radiation}

%

\author{Yoshiyuki Inoue}
\affiliation{JAXA International Top Young Fellow, Institute of Space and Astronautical Science JAXA, 3-1-1 Yoshinodai, Chuo-ku, Sagamihara, Kanagawa 252-5210, Japan}
\email{yinoue@astro.isas.jaxa.jp}

\begin{abstract}
The cosmic gamma-ray background radiation is one of the most fundamental observables in the gamma-ray band. Although the origin of the cosmic gamma-ray background radiation has been a mystery for a long time, the {\it Fermi} gamma-ray space telescope has recently measured it at 0.1--820~GeV and revealed that the cosmic GeV gamma-ray background is composed of blazars, radio galaxies, and star-forming galaxies. However, {\it Fermi} still leaves the following questions. Those are dark matter contribution, origins of the cosmic MeV gamma-ray background, and the connection to the IceCube TeV--PeV neutrino events. In this proceeding, I will review the current understandings of the cosmic gamma-ray background and discuss future prospects of cosmic gamma-ray background radiation studies. I also briefly review the current status of cosmic infrared/optical background radiation studies.
\end{abstract}

\maketitle

\thispagestyle{fancy}


\section{Introduction}
\begin{figure}[t]
\centering
\includegraphics[bb = 0 0 360 252, width=85mm]{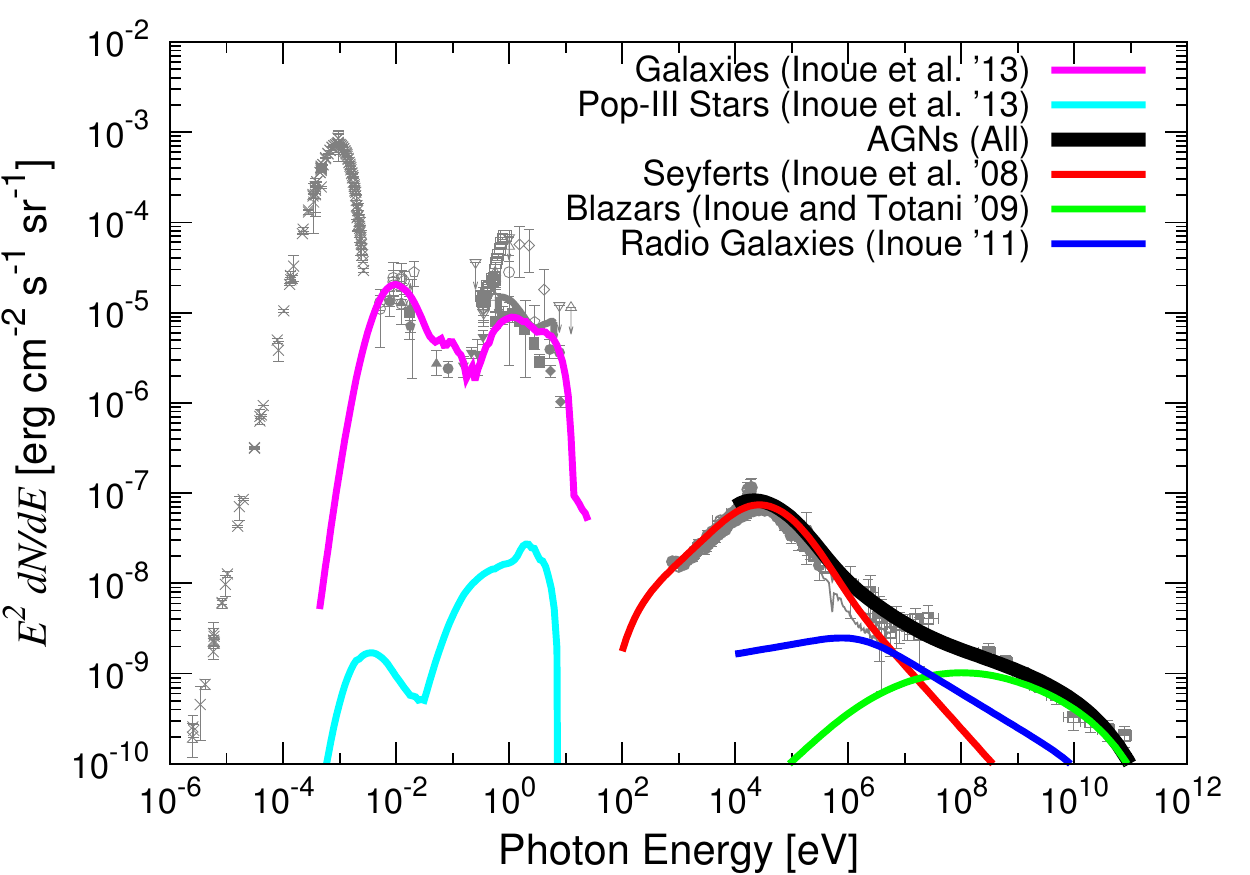}
\caption{The cosmic background radiation spectrum from microwave to gamma-ray energies. Contribution from galaxies \cite{ino13_cib}, Pop-III stars \cite{ino13_cib}, Seyferts \cite{ino08}, blazars \cite{ino09}, radio galaxies \cite{ino11}, and all AGNs is shown by purple, cyan, red, green, blue, and black curve, respectively. The references for the measurements are in \cite{ino13_cib,ino13_cxb}.} \label{fig:cb_all}
\end{figure}

The cosmic background radiation is one of the most fundamental observables from the sky. It is the result of integrated emission from its origins over the cosmic history. Figure. \ref{fig:cb_all} shows the measured cosmic background radiation spectrum from microwave to gamma rays. 

The origins of cosmic background radiation from microwave to X-ray are well understood. For example, the cosmic X-ray background (CXB) has been conclusively shown to be the integrated light produced via the accretion process of active galactic nuclei (AGNs), in particular Seyferts, hosting supermassive black holes \cite[e.g.][]{ued14}. By contrast, the origin of the cosmic gamma-ray background (CGB)\footnote{The cosmic gamma-ray background (CGB) is also called as the extragalactic gamma-ray background (EGRB or EGB) or the isotropic gamma-ray background (IGRB).} has been an intriguing mystery for these forty years since its discovery by the {\it SAS}-2 satellite \cite{fic77,fic78}. Moreover, gamma-ray signatures from dark matter particles are expected to be buried in the CGB. The CGB has drawn a lot of attention from the community for a long time.

Before the {\it Fermi} gamma-ray space telescope (hereinafter {\it Fermi}) era, neither spectrum nor origins of the cosmic GeV gamma-ray background were not well understood. Although EGRET onboard the {\it CGRO} satellite reported the spectrum at 0.03--50~GeV \cite{sre98,str04}, an anomaly was known to exist at GeV energies, the so-called EGRET GeV anomaly \cite{ste08}. Regarding the origins of the background, blazars were expected to explain it. However, due to small EGRET samples, it was predicted that blazars' contribution is at the level of 20-100\% depending on models \cite[see e.g.][and references therein]{ino09}. More accurate determination of the CGB spectrum and more extragalactic source samples were required to understand the nature of the CGB. {\it Fermi} has recently reported a broadband CGB spectrum and the large {\it Fermi} source catalog has enabled us to unveil the origins of the CGB at the GeV gamma-ray band. At the same time, the current {\it Fermi} measurement still leaves the following unanswered problems. First, the signature of dark matter annihilation/decay has not yet observed in the CGB. Second, the origins of the cosmic MeV gamma-ray background are not understood at all. Lastly, the cosmic TeV gamma-ray background has not been explored yet. Especially understanding of its connection to the recent IceCube neutrino events will be an important key in this multi-messenger astronomy era. 

In this paper, I review our current understandings of the cosmic GeV gamma-ray background radiation in \S.\ref{sec:cgb_gev}. Then, future prospects of the cosmic gamma-ray background studies will be discussed in \S.\ref{sec:prospects}. I also briefly review the current status of cosmic infrared/optical background radiation studies in \S.\ref{sec:cib}. Summary is given in \S.\ref{sec:sum}.

\section{Cosmic GeV Gamma-ray Background Radiation}
\label{sec:cgb_gev}

\subsection{Measurements}

{\it Fermi} has recently allowed a broadband and accurate measurement of the CGB spectrum between 0.1--820~GeV \cite{ack14_cgb}, which is updated from the previous report \cite{abd10_cgb}. The anomaly seen in the EGRET CGB  spectrum disappeared. {\it Fermi} has resolved $\sim30$\% of the cosmic GeV gamma-ray background to point sources at $\sim1$~GeV and resolved more at higher energies. This implies that current and future Cherenkov gamma-ray telescopes will be able to reveal a great portion of the CGB at $\gtrsim100$~GeV with their better sensitivities at these energies \cite{fun13}. The resulting unresolved background spectrum is found to be compatible with a power law with a photon index of 2.32 that is exponentially cut off at 279~GeV. The total intensity of the unresolved CGB is $7.2\times10^{-6}\ \rm{cm^{-2}s^{-1}sr^{-1}}$ above 0.1~GeV. The measured cut-off signature may reflect gamma-ray attenuation by the cosmic infrared/optical background. However, further studies will be required to conclude it as the result of the gamma-ray attenuation, since it also depends on intrinsic spectra and evolution of source populations.

\begin{figure}[t]
\centering
\includegraphics[bb = 822 365 255 20, width=85mm]{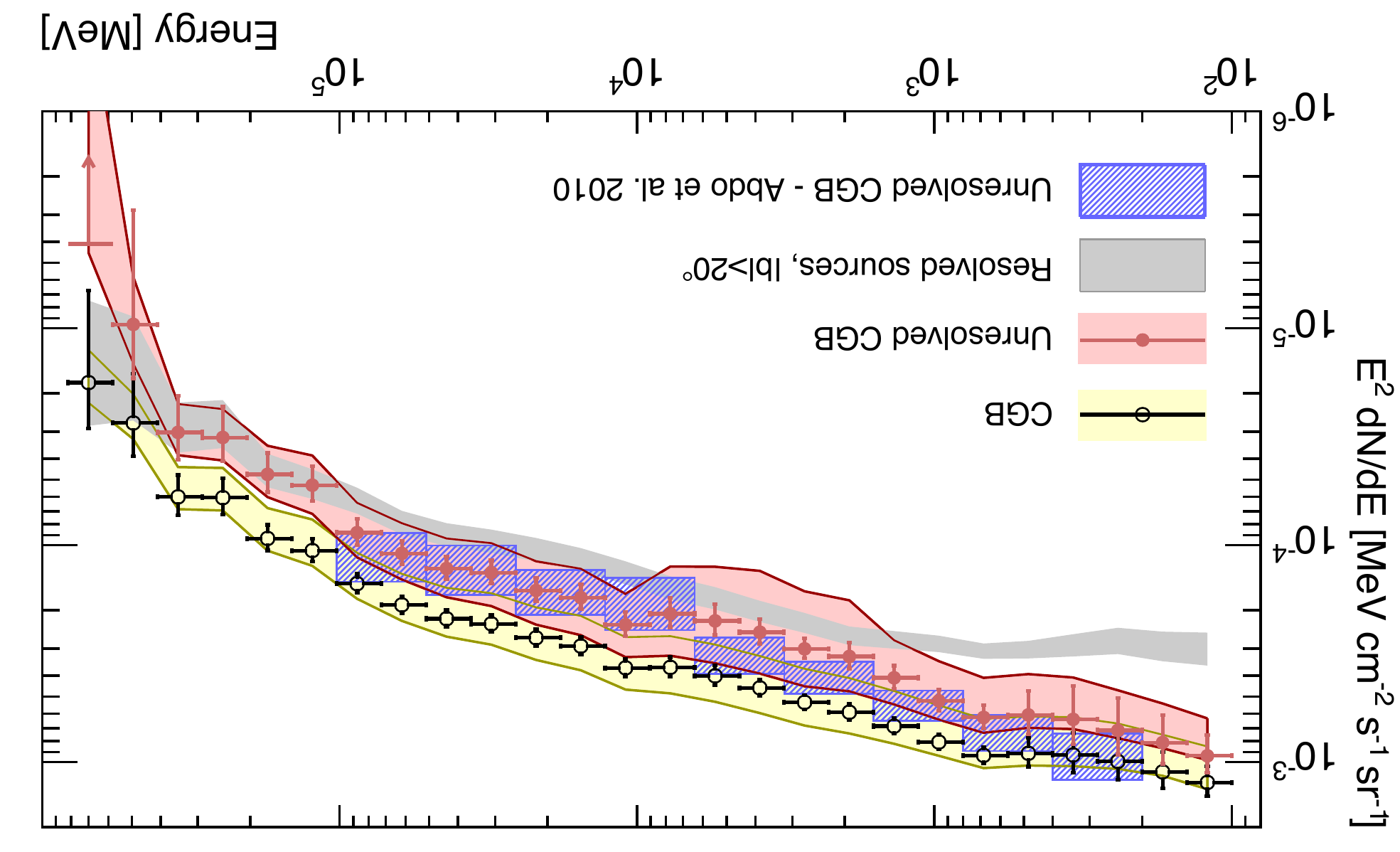}
\caption{The CGB intensities measured by {\it Fermi}. The error bars include the statistical uncertainty and systematic uncertainties. The shaded  bands indicate the systematic uncertainty arising from uncertainties in the Galactic foreground. The CGB intensity is the sum of the unresolved CGB and the resolved LAT sources (resolved CGB) at high Galactic latitudes, $|b|>20\deg$. Taken from Ackermann et al. 2014 \cite{ack14_cgb}, but the legends in the plot are modified.}
\label{fig:cgb_fermi}
\end{figure}

\subsection{Compositions}
Various gamma-ray emitting sources have been discussed as the origins of the cosmic GeV gamma-ray background in the literature. Those are blazars \cite[e.g.][]{pad93,ste93,ino09}, star-forming galaxies \cite[e.g.][]{str76,pav02}, radio galaxies \cite[e.g.][]{str76,pad93,ino11}, gamma-ray bursts (GRBs) \cite[e.g.][]{cas07}, high Galactic-latitude pulsars \cite[e.g.][]{fau10}, intergalactic shocks \cite[e.g.][]{loe00,tot00}, Seyferts \cite[e.g.][]{ino08}, cascade from ultra-high-energy cosmic rays \cite[e.g.][]{dar95,kal09}, large galactic electron halo \cite{kes04}, cosmic-ray interaction in the solar system \cite{mos09}, and dark matter annihilation/decay \citep[e.g.][]{ber01}. Among these possible candidates, {\it Fermi} has observed gamma rays from blazars, star-forming galaxies, radio galaxies, GRBs, and high-latitude pulsars \citep{nol12}. These are guaranteed populations contributing to the CGB. Since gamma-ray bursts and high-latitude pulsars are known to make little contribution \cite{cas07,sie11}, I focus on blazars, radio galaxies, and star-forming galaxies below.

\begin{figure*}[t]
\centering
\includegraphics[bb = 0 0 624 453, width=120mm]{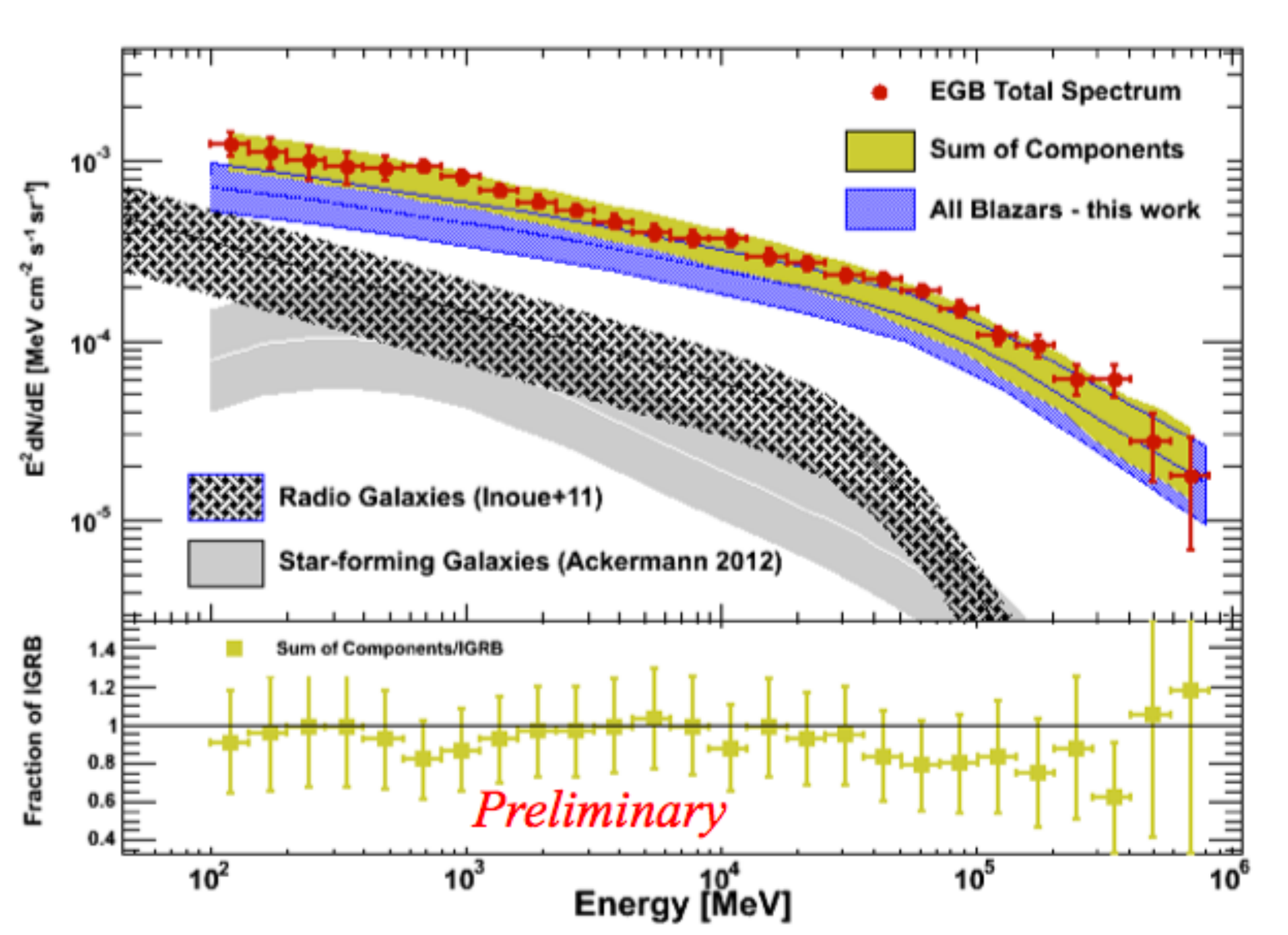}
\caption{Top Panel: The cosmic gamma-ray background spectrum of blazars (bule \cite{aje14}), radio-galaxies (black striped, \cite{ino11}), star-forming galaxies, and summation of these three populations (yellow). The CGB measurement is shown by red points. Bottom Panel: the residual emission, computed as the ratio of the summed contribution to the CGB spectrum. Taken from Ajello et al. 2014 presented at the workshop "High Energy Messengers: Connecting the Non-Thermal Extragalactic Backgrounds" \cite{aje14_chicago}.}
\label{fig:cgb_comp}
\end{figure*}

\subsubsection{Blazars}
Blazars emit gamma rays via the inverse Compton scattering processes (e.g. \cite{ree67,mar92,der93,sik94}, but see also hadronic processes \cite{man89,muc03}). Observationally, blazars are known to be divided into two population, flat spectrum radio quasars (FSRQs) and BL Lacs, and it has been suggested that the spectral energy distributions (SEDs) of blazars evolve with luminosity, as described by the so-called blazar sequence \cite{fos98,kub98}. They are dominant extragalactic gamma-ray sources \cite{har99,nol12}. Therefore, it is naturally expected that blazars explain the cosmic GeV gamma-ray background \cite[see e.g.][]{ino09}. However, its fraction was very uncertain in the EGRET era due to its small samples. 

With the large sample of gamma-ray blazars, {\it Fermi} has enabled us to accurately evaluate the cosmological evolution of blazars. Ajello et al. (2012) \cite{aje12} constructed the gamma-ray luminosity function (GLF) of FSRQs with the {\it Fermi} FSRQ samples. Regarding BL Lacs, redshifts of  about half of BL Lacs are not determined, which makes difficult to construct GLFs of BL Lacs. Recently Ajello et al. (2014) \cite{aje14} have successfully constructed the GLF of BL Lacs by using redshift constraints on individual BL Lacs. These studies confirmed that FSRQs and BL Lacs, i.e. blazars, show the luminosity-dependent density evolution like X-ray AGNs \cite{ued14}, which was suggested since the EGRET era \cite{nar06,ino09}. Base on these efforts, blazars are known to explain $\sim$50\% of the CGB flux above 0.1~GeV \cite[e.g.][]{aje14_chicago}. At higher energies ($\gtrsim100$~GeV), blazars explain $\sim$100\% of the cosmic gamma-ray background flux. 

Interestingly, HBLs (low-luminosity BL Lacs) show strong negative cosmological evolution, while FSRQs and luminous BL Lacs (so-called LBLs and IBLs) show positive evolution like Seyferts and the cosmic star formation history \cite{ued14,hop06}. From other wavelength studies, BL Lacs were known to show no or negative evolution  (e.g. \cite{gio99,pad07} but see also \cite{mar13} reporting positive evolution). Such negative evolution is different from the evolution of other AGN populations. The understanding of the physical cause of this negative evolution may be an important key to understand the AGN evolutionary history. 

\subsubsection{Radio Galaxies}
{\it Fermi} has detected gamma rays from $\sim$10 misaligned AGNs (i.e., radio galaxies) \cite{abd10_magn}. Although they are fainter than blazars, radio galaxies in the entire sky are more numerous than blazars. It is naturally expected that they will make a significant contribution to the CGB. To study the contribution of gamma-ray-loud radio galaxies to the CGB, their GLF is required. However, it is not straightforward to construct it because of the limited {\it Fermi} radio galaxy samples. On the contrary, the radio luminosity function (RLF) of radio galaxies is well established \cite[e.g.][]{wil01}, since they are mainly detected in the radio band. Therefore, by using the correlation between radio and gamma-ray luminosities of radio galaxies, we are able to convert the RLF to the GLF. Base on this method, radio galaxies are expected to explain $\sim20$~\% of the CGB at $>0.1$~GeV \cite{ino11,dim14}. 

\subsubsection{Star-forming Galaxies}
{\it Fermi} has also detected gamma rays from 7 star-forming galaxies \cite{abd10_magn}. Those gamma rays are produced interactions of cosmic rays with gas or interstellar radiation fields. Since there are numerous galaxies in the sky, they are also expected to contribute to the CGB. However, similar to radio galaxies, GLF of star-forming galaxies can not be constructed solely with the {\it Fermi} samples. Therefore, using the correlation between infrared and gamma-ray luminosities, both of which trace the star formation activity, the contribution of star-forming galaxies to the background can be estimated with the well-established infrared luminosity functions \cite[e.g.][]{pav02,ack12}. The expected contribution is 10--30\% of the CGB at $>0.1$~GeV \cite{ack12}. 

Figure. \ref{fig:cgb_comp} show the contribution of these three populations to the cosmic gamma-ray background spectrum. Summing blazars, radio galaxies, and star-forming galaxies, we can explain $\sim90$\% of the CGB radiation at $>0.1$~GeV. By considering the measurement and model uncertainties, {\it Fermi} has enabled us to unveil that the cosmic GeV gamma-ray background is composed of blazars, radio galaxies, and star-forming galaxies. However, I note that radio galaxies and star-forming galaxies still rely on the luminosity correlation due to the small samples. This situation is very similar to blazar studies in the early EGRET era \cite{pad93,ste93,ste96}. Therefore, further data will be required to precisely evaluate the fraction of these two populations.

%



\section{Prospects for future Cosmic Gamma-ray Background Radiation Studies}
\label{sec:prospects}
{\it Fermi} has unveiled the origins of the cosmic GeV gamma-ray background. However, there are still three unsolved questions. First, we do not see the signature of dark matter particles in the CGB spectrum, although they are expected to contribute to the CGB. How can we probe dark matters with future cosmic gamma-ray background studies? Second, although the cosmic X-ray and GeV gamma-ray backgrounds are well understood, the cosmic MeV background has not been fully explored yet due to the observational difficulties. What are the origins of the cosmic MeV gamma-ray background? Lastly, the cosmic TeV gamma-ray background has never been observed yet, although {\it Fermi} has revealed the CGB up to 820~GeV. Recently, IceCube detected a few tens of TeV--PeV neutrino events. Although their origins are still debated, those neutrino events should be associated with gamma rays at those energies. How are the cosmic gamma-ray background and the cosmic TeV--PeV neutrino background connected? I briefly summarize future prospects of these three issues.

\subsection{{\it Anisotropy} of the Cosmic GeV Gamma-ray Background Radiation}
{\it Fermi} has accurately measured the cosmic GeV gamma-ray background spectrum \cite{ack14_cgb}. {\it Fermi} has also measured the anisotropy, i.e. the angular power spectrum, of the CGB at 1--50~GeV \cite{ack12_anis}. Ando \& Komatsu (2006) \cite{and06} proposed that anisotropy in the CGB is a smoking-gun signature of annihilation of dark matter particles. Since dark matter traces the large-scale structure of the universe, the emission from dark matter is anisotropic and its spatial pattern is unique and predictable \cite[e.g.][]{and07}. The measured angular power spectrum was consistent with the constant value at all multipoles, which means the Poisson term, so-called the shot-noise, dominates the signals. By comparing the expected CGB angular power spectrum from dark matter annihilation with the measured power spectrum, upper limits on the annihilation crosse section are placed \cite{and13}. The current data exclude $<\sigma v>\gtrsim10^{-25}\ {\rm cm^3s^{-1}}$ for  annihilation into $bb^-$ at the dark matter mass of 10~GeV. Since the analysis was based on the first 22-month data, the limits can be improved further with current and future {\it Fermi} data.

\begin{figure}[t]
\centering
\includegraphics[bb = 0 0 531 500, width=70mm]{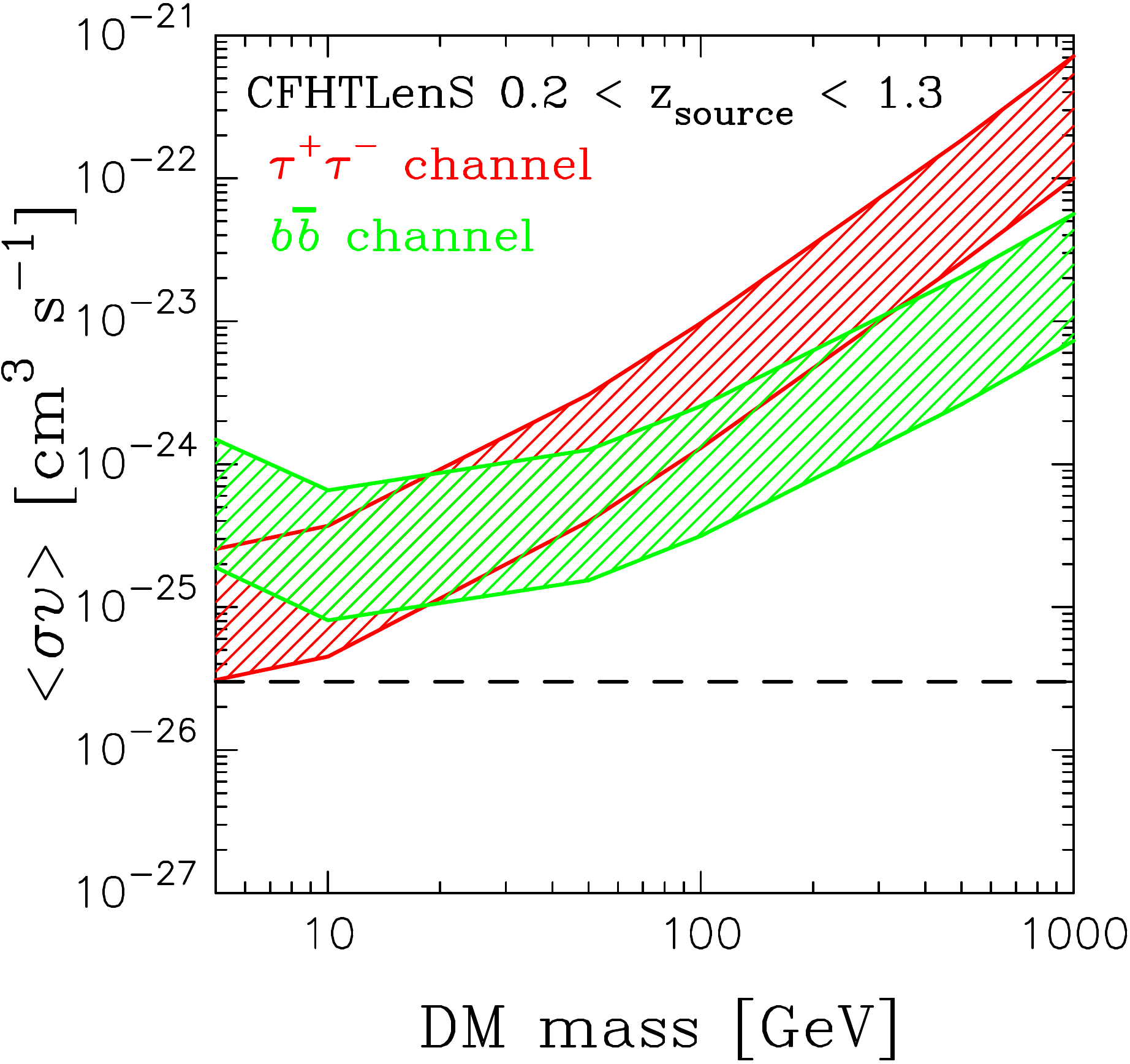}
\caption{ The 68 \% confidence level upper limits on $\langle \sigma v \rangle$ as a function of DM mass using the cross correlation of cosmic shear and the CGB emission. The cosmic shear data from the CFHTLenS (154~deg$^2$) is used here. The red shaded region shows the upper bound for the $\tau^{+}\tau^{-}$  channel and the green region is for the $b {\bar b}$ channel. Note that the widths of the shaded regions indicate the model uncertainty. Taken from Shirasaki et al. 2014 \cite{shi14}.} \label{fig:cgb_ani_DM}
\end{figure}

Sirasaki et al. (2014) \cite{shi14} have recently reported the first measurement of the cross correlation of weak gravitational lensing and the CGB emission (see Figure. \ref{fig:cgb_ani_DM}). The cross correlation is also a powerful probe of signatures of dark matter annihilation \cite{cam13}, because both cosmic shear and gamma-ray emission originate directly from the same dark matter distribution in the Universe. Using the cosmic shear data from the CFHTLenS (154~deg$^2$), they exclude dark matter annihilation cross sections of $<\sigma v>=10^{-24}$--$10^{-25}\ {\rm cm^3s^{-1}}$ for 100~GeV dark matter. There are several wider optical survey projects in near future such as the Subaru Hyper Suprime-Cam (HSC), the Dark Energy Survey, and the Large Synoptic Survey Telescope (LSST). HSC and LSST will cover 1400~deg$^2$ and 20000~deg$^2$, respectively. These future surveys will increase the sensitivity to probe the dark matter annihilation cross sections.

The anisotropy is also useful to constrain the cosmological evolution of sources. Since the number density of blazars are smaller than radio galaxies and star-forming galaxies, blazars contribute to the anisotropy more significantly than the other two do. Therefore, the anisotropy measurement enabled us to put constraints on the blazar evolution models \cite{cuo12,har12,dim14_anis}. 

\subsection{Cosmic {\it MeV} Gamma-ray Background Radiation}
By contrast to the cosmic X-ray/GeV gamma-ray backgrounds, the origin of the cosmic MeV gamma-ray background at $\sim1-10$ MeV is still an intriguing mystery. The background spectrum from several hundreds keV to several tens MeV is smoothly connected to the CXB spectrum and shows much softer than the GeV component \cite{fuk75,wat97,wei00}, indicating a different origin from that above 100 MeV.  The Seyfert spectra adopted in population synthesis models of the CXB cannot explain this component because of the assumed exponential cutoff at a few hundred keV, where thermal hot corona above the accretion disk is assumed. 

Several candidates have been proposed to explain the MeV background. One was the nuclear-decay gamma-rays from Type Ia supernovae \cite[SNe Ia;][]{cla75}. However, recent measurements of the cosmic SN Ia rates show that the expected background flux is about an order of magnitude lower than observed \cite{ahn05_sn,hor10}. Seyferts can naturally explain the MeV background including the smooth connection to the CXB \cite{sch78,ino08}. Comptonized photons produced by non-thermal electrons in coronae surrounding accretion disks can produce the MeV power-law tail \cite{ino08}. There is also a class of blazars, called MeV blazars which are FSRQs, whose spectra peak at MeV energies. These MeV blazars could potentially contribute to the MeV background as well \cite{aje09}. Radio galaxies have been also discussed as the origin of the MeV background \cite{str76}. However, recent studies show that the expected background flux from radio galaxies is $\sim10$\% of the total MeV background flux \cite{mas11,ino11}. Annihilation of the dark matter particles has also been discussed \cite{oli85,ahn05_dm1,ahn05_dm2}, but those are less "natural" dark matter candidates, with a mass scale of MeV energies, rather than GeV-TeV dark matter candidates. In either cases, there is little observational evidence of MeV emission from these candidates and a quantitative estimate is not easy due to the sensitivity of the MeV measurements.

\begin{figure}[t]
\centering
\includegraphics[bb = 0 0 360 252, width=80mm]{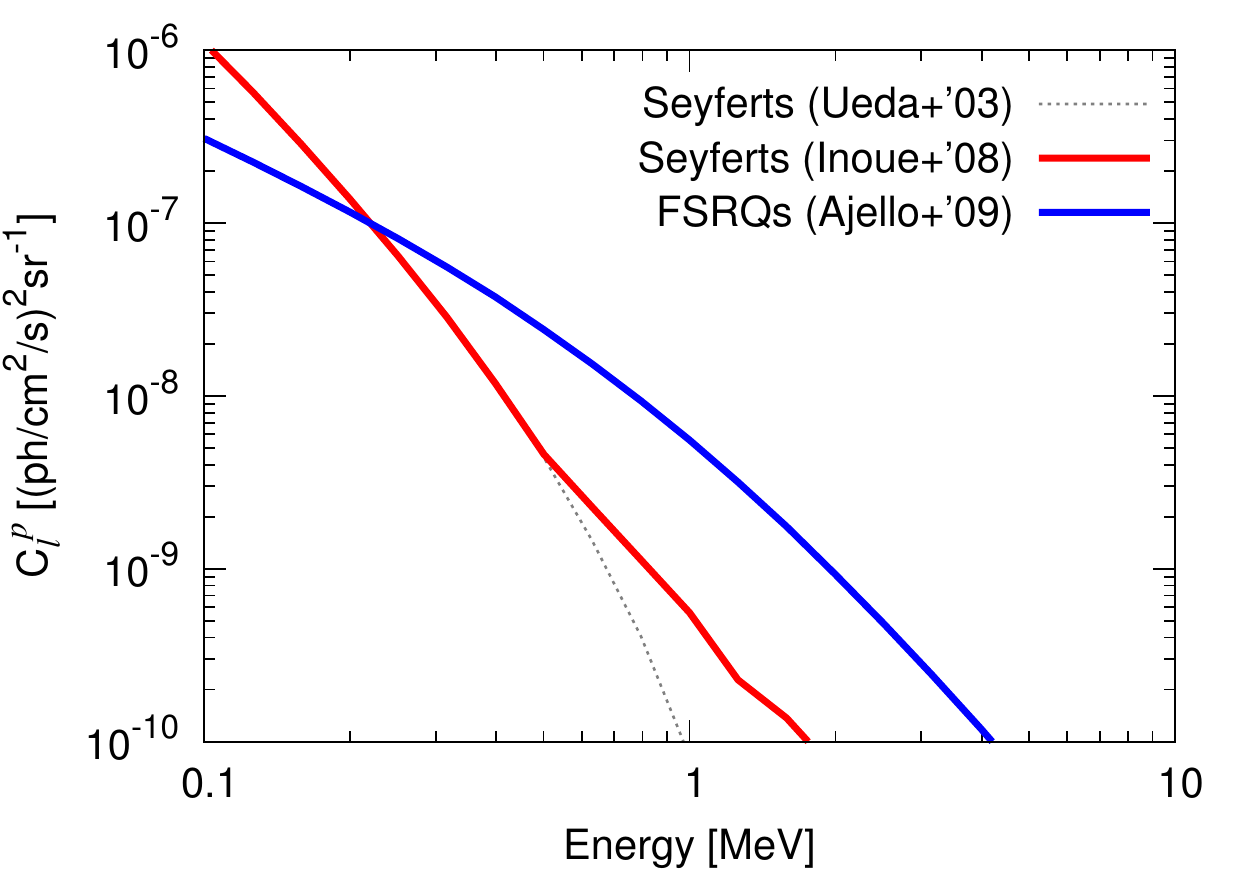}
\caption{Expected Poisson term of the angular power spectrum of the cosmic MeV gamma-ray background at 200 keV-- 10 MeV. Red and blue curve corresponds to Seyferts with non-thermal electrons in coronae \cite{ino08} and FSRQs \cite{aje09}, respectively, assuming the MeV background is explained by them. For reference, we also plot the model of Seyferts with thermal cutoff \cite{ued03} by dotted curve which does not explain the MeV background radiation. Taken from Inoue et al. 2013 \cite{ino13_cxb}} \label{fig:aniso}
\end{figure}

It is not easy to resolve the MeV sky as in the soft X-ray or the GeV gamma-ray bands. However, anisotropy in the background radiation will shed new light on this problem as in the GeV gamma-ray background \cite{and06}. Fig. \ref{fig:aniso} shows the Poisson term of the angular power spectra of Seyferts with non-thermal components in coronae \cite{ino08} and FSRQs \cite{aje09}. For reference, we also plot Seyferts with simple thermal cutoff spectra \cite{ued03}, but note that those do not explain the MeV background. Since the contribution of the correlation term is negligible in this energy region and the assumed flux limits, the angular power spectrum is dominated by the Poisson term. This Poisson term measurement is useful enough to distinguish the origin of the MeV background.  The difference of the $C_l^p$ of Seyferts \cite{ino08} and FSRQs \cite{aje09} is more than an order of magnitude. The reason why we can clearly distinguish the origin is as follows. Seyferts are fainter but more numerous than blazars. These two differences are able to make future MeV instruments such as the SGD on board {\it Astro-H} \cite{tak12} clearly detect the origin of the MeV gamma-ray sky through the angular power spectrum of the sky. 
  
If the origin of the MeV background is non-thermal emission from Seyfert \cite{ino08}, this implies that magnetic reconnection heats the corona above the disk and accelerate non-thermal electrons in the corona. As discussed in \cite{ino08,ino14_corona}, this scenario will be also tested by future X-ray and sub-mm observations of individual sources. If it is FSRQs \cite{aje09}, this implies that there are two distinct FSRQ populations in MeV and GeV because of the spectral difference between the MeV and GeV background. This will suggest that there are two different populations in FSRQs between MeV and GeV. This may pose a problem to the AGN unification scheme.

\subsection{Cosmic {\it TeV} Gamma-ray Background Radiation}
\begin{figure}[t]
\centering
\includegraphics[bb = 0 0 360 252, width=80mm]{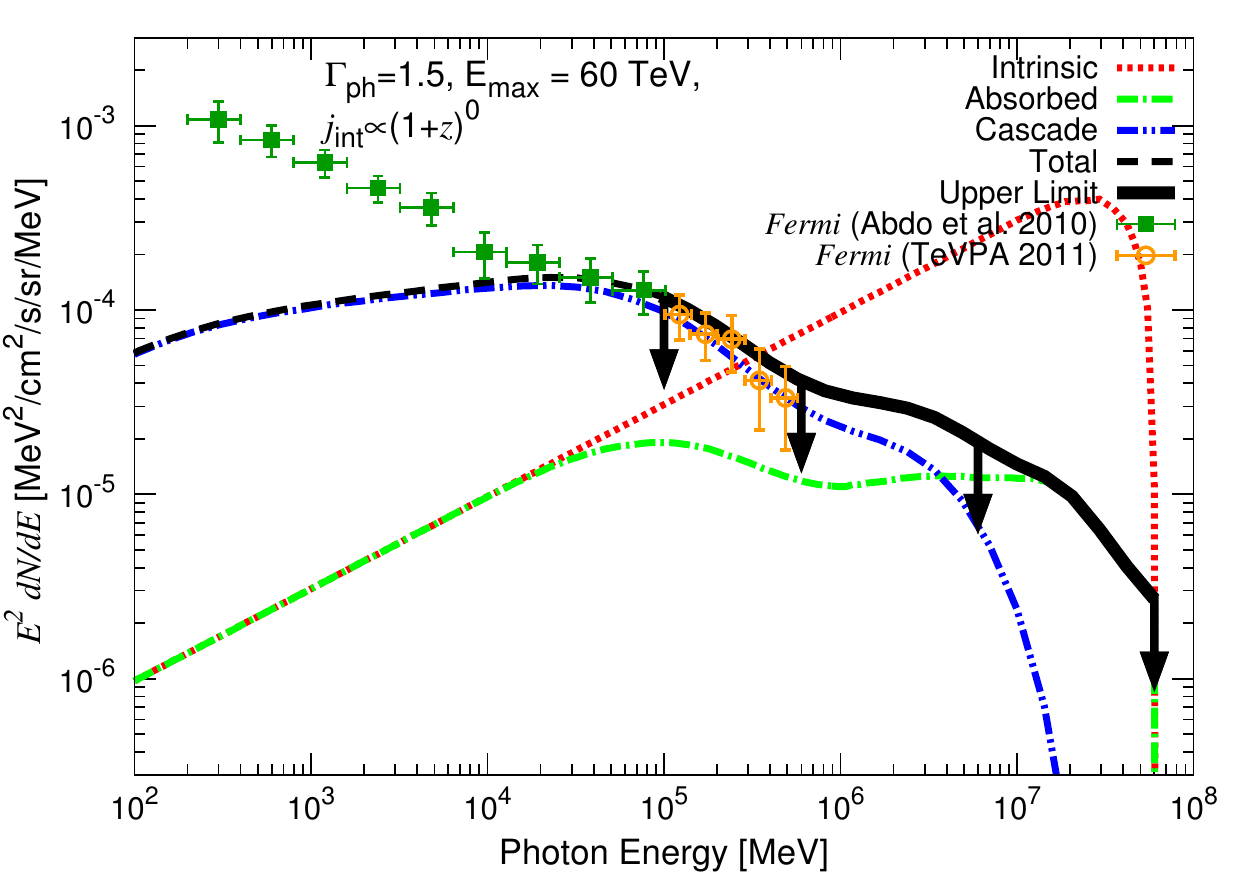}
\caption{Upper limit on the CGB by requiring the cascade emission not to exceed the CGB data below 100~GeV in the model-independent way. We set the photon index $\Gamma=1.5$ and the maximum energy $E_{\rm max}=60$~TeV with no redshift evolution. Dotted, dot--dashed, double dot--dashed and dashed curves show the intrinsic spectrum (no absorption), absorbed, cascade, and total (absorbed+cascade) CGB spectrum, respectively. Thick solid curve with arrows show the upper limit.  The filled square points show the observed CGB data with the 11-months {\it Fermi} data \cite{abd10_cgb}. The circle points show the observed CGB data with the 24-months {\it Fermi} data \cite{ack11_TeVPA}. Error bars represent 1-$\sigma$ uncertainty of the data. Taken from Inoue \& Ioka 2012 \cite{ino12}} \label{fig:cgb_ul}
\end{figure}

The cosmic TeV gamma-ray background has not been observed yet, although {\it Fermi} has measured the CGB up to 820~GeV. Here, very high energy (VHE; $\gtrsim100 {\rm GeV}$)  gamma-rays propagating through the universe experience absorption by the interaction with the cosmic optical/infrared background (COB and CIB) via electron--positron pair production \cite[e.g.][]{ino13_cib}. Those generated electron--positron pairs scatter the cosmic microwave background (CMB) radiation via the inverse Compton scattering  and generate secondary gamma-ray emission component (the so-called cascade emission) in addition to the absorbed primary emission \footnote{The pairs may loose their energy through the plasma beam instabilities (\cite{bro12,cha14}, but see also \cite{sir14}).}. At redshift $z$, the scattered photon energy $E_{\gamma,c}$ appears at lower energy than the intrinsic photon energy $E_{\gamma,i}$, typically $E_{\gamma,c}\approx 0.8\,(1+z)\left(E_{\gamma,i}/{1\,{\rm TeV}}\right)^2 \  {\rm GeV}.$ This cascade component is also expected to contribute to the CGB  \cite{cop97,mur12, ino12}. Therefore, the current CGB measurement below 100 GeV sets an upper limit on the CGB itself at the TeV gamma-ray band. The limit is conservative for the electromagnetic cascade emission from the VHE CGB interacting with the cosmic microwave-to-optical background radiation {\it not} to exceed the current CGB measurement \cite{ino12} (See Figure \ref{fig:cgb_ul}).

\begin{figure}[t]
\centering
\includegraphics[bb = 0 0 216 151, width=80mm]{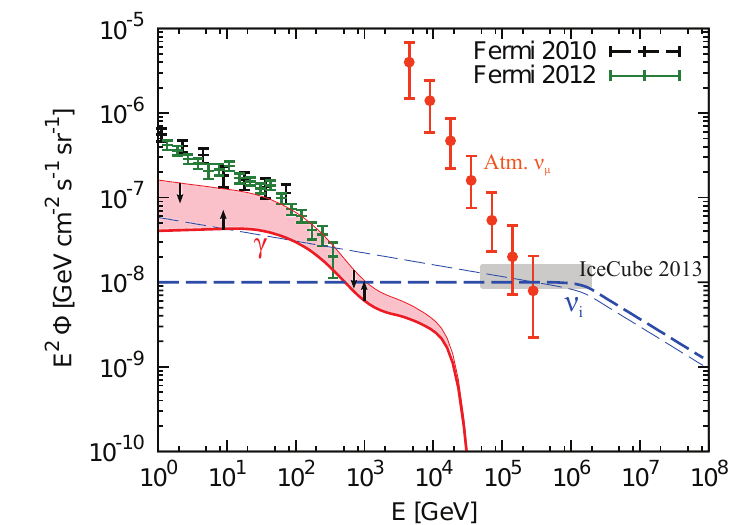}
\caption{The allowed range in $pp$ scenarios explaining the measured IceCube neutrino flux, which is indicated by the shaded area with arrows. The integrated neutrino background (dashed) and corresponding CGB (solid) are shown for $\Gamma=2.0$ (thick) and $\Gamma=2.14$ (thin). No redshift evolution is assumed here. The shaded rectangle indicates the IceCube data \cite{aar13}. Taken from Murase et al. 2013 \cite{mur13}.} \label{fig:inb_mur13}
\end{figure}

Measurements of the cosmic TeV gamma-ray background spectrum is also important from the multi-messenger point of view. The IceCube Collaboration has recently reported the detection of TeV--PeV neutrinos for the first time \cite{aar13,aar14}. This detection opens up a multi-messenger connection among photons, neutrinos, and cosmic rays. The origin of the IceCube neutrinos are still under debate \cite[see ][for reviews]{mur14}. Conventionally, those high energy neutrinos are produced by cosmic rays via hadronuclear ($pp$) and/or photohadronic ($p\gamma$) interactions. In either cases, gamma rays must be produced. If the IceCube events originate in the extragalactic sky, the origins of these neutrino events are also responsible for the cosmic TeV gamma-ray background radiation. Importantly, the measured cosmic gamma-ray background flux around 0.7~TeV is $E_\gamma^2dN_\gamma/dE_\gamma\sim2\times10^{-8}\ {\rm GeV\ cm^{-2}\ s^{-1}\ sr^{-1}}$, while the measured neutrino background flux in the 100~TeV--PeV range is $E_\nu^2dN_\nu/dE_\nu\sim10^{-8}\ {\rm GeV\ cm^{-2}\ s^{-1}\ sr^{-1}}$ per flavor \cite{aar13,aar14}.

In $p\gamma$ scenarios, secondary spectra have a strong dependence on target photon field energy distribution \cite{mur14_seq,der14}, while $pp$ scenarios give power-law secondary spectra following the initial cosmic-ray spectrum. Therefore, the current {\it Fermi} and IceCube measurements put powerful constraints on $pp$ scenarios \cite{mur13}. The allowed range in $pp$ scenarios explaining the neutrino events is shown in Figure. \ref{fig:inb_mur13}, taken from Murase, Ahlers, and Lacki (2013) \cite{mur13}, in which the gamma-ray attenuation by the CIB/COB is taken into account. The IceCube data indicate that these sources contribute at least 30--40\% of the cosmic very high energy gamma-ray background and even $\sim100$\% for softer spectra. As discussed above, blazars, which are not $pp$ sources, are responsible for $\sim100$\% of that background flux, which is inconsistent with the discussion here. Therefore, further studies in these fields including evolution and SED of blazars are required to ease this tension.

\section{Cosmic Optical and Infrared Background Radiation}
\label{sec:cib}

The determination of the cosmic optical and infrared background radiation, sometimes called the extragalactic background light (EBL), is important to the CGB science. The COB/CIB, the diffuse, isotropic background radiation from far-infrared (FIR) to ultraviolet (UV) wavelengths, is believed to be predominantly composed of the light from stars and dust integrated over the entire history of the universe \cite[see][for reviews]{dwe13}. The observed spectrum of the local COB/CIB at $z = 0$ has two peaks of comparable energy density. The first peak in the optical to the near-infrared (NIR) is attributed to direct starlight, while the second peak in the FIR is attributed to emission from dust that absorbs and reprocesses the starlight \cite{dom11,ino13_cib}. The energy density of the local COB/CIB has been constrained to be $<$ 24 nW m$^{-2}$ sr$^{-1}$ at optical wavelengths, and $<$ 5  nW m$^{-2}$ sr$^{-1}$ between 8 $\mu$m and 31 $\mu$m \cite{mey12}. Combined with the lower limits from galaxy counts, the total EBL intensity at $z=0$ from 0.1 $\mu$m to 1000 $\mu$m is inferred to lie in the range 52--99 nW m$^{-2}$ sr$^{-1}$ \cite{hor09}. 

Integration over galaxy number counts provides a firm lower bound on the COB, and the observed trend of the counts with magnitude indicates that the COB at $z = 0$ has been largely resolved into discrete sources in the optical/NIR bands \cite[see e.g.][]{mad00,tot01}. However, the precise determination of the COB/CIB has been a difficult task. Direct measurements of the COB/CIB in the optical and NIR bands have been hampered by bright foreground emission caused by interplanetary dust, the so-called zodiacal light. An excess from galaxy counts in the NIR background has been reported by several experiments \cite[see e.g.][]{mat05,tsu13}. Although this excess can be explained by redshifted light from first stars \cite[][]{mat05}, reionization observations disfavor such scenario which leads overproduction of ionizing photons at high redshifts \cite{mad05,ino14}. Later it was also found that this excess would be inconsistent with TeV observations of nearby blazars (e.g. \cite{aha06}, but see also \cite{ess10}). Recently, Matsuoka et al. (2011) \cite{mat11} reported measurements of the COB at 0.44 $\mu$m and 0.65 $\mu$m from outside the zodiacal region using observational data from Pioneer 10/11, which are consistent with the galaxy counts. 

The COB/CIB can also be probed indirectly through observations of high-energy gamma rays from extragalactic objects \cite[e.g.][]{gou66,jel66,ste92}. Gamma rays propagating through the intergalactic space are attenuated by the pair production interactions with low-energy photons of the COB/CIB. For gamma-rays of given energy $E_\gamma$, the pair production cross section peaks for low-energy photons with energy $\epsilon_{\rm peak}\simeq 2m_e^2c^4/E_\gamma\simeq0.5\left(1{\rm \ TeV}/E_\gamma\right)\ {\rm eV}$, where $m_e$ is the electron mass and $c$ is the speed of light. In terms of wavelength, $\lambda_{\rm peak} \simeq2.5(E_\gamma[{\rm TeV}])\ \mu{\rm m}$. Measuring the attenuation features in the spectra of extragalactic GeV-TeV sources offers a valuable probe of the COB/CIB that is indirect. Although this method can be limited by incomplete knowledge of the intrinsic spectra of the source before attenuation, by assuming a plausible range for such spectra, observations of blazars by current ground-based telescopes have been able to place relatively robust upper limits to the COB/CIB \cite[e.g.][]{aha06}. This has been complemented by {\it Fermi} observations of blazars and GRBs that placed upper limits on the $\gamma\gamma$ opacity up to $z=4.35$ \cite{abd09_080916C}. Very recently, H.E.S.S. has successfully measured the imprint of the local COB/CIB in the spectra of bright blazars, assuming only that their intrinsic spectra have smooth shapes \cite{abr13}. {\it Fermi} has also positively detected the redshift-dependent signature of the COB attenuation up to $z=1.5$, utilizing the collective spectra of a large number of blazars \cite{ack12_ebl}. However, recently it has also known that there is a class of extreme HBLs which shows very hard spectra at the TeV gamma-ray band \cite[see e.g.][]{ino13_cib}. As their emission mechanism is still under debate \cite{ess10,lef11}, further careful analysis on the COB/CIB determination from gamma-ray observations would be required.

Interestingly, very recently the CIBER collaboration has reported an excess in the CIB fluctuation from galaxies' contribution at 1.1 and 1.6 $\mu$m \cite{zem14}, which has previously reported at 3--5~$\mu$m by {\it Spitzer} and {\it AKARI} \cite{kas05,coo07,mat11_flu}. This discovery may suggest a new population in the CIB other than galaxies. Further studies such as the CIB spectrum measurements will be important to unveil the origin of this excess. 

\section{Summary}
\label{sec:sum}
{\it Fermi} has very recently allowed a broadband measurement of the cosmic GeV gamma-ray background spectrum between 0.1--820~GeV. {\it Fermi} has resolved $\sim$30\% of it to point sources at $\sim$1~GeV and more fraction at higher energies. The unresolved background spectrum is compatible with a power law with a photon index of 2.32 that is exponentially cut off at 279~GeV. Moreover, theoretical works based on the {\it Fermi} measurements have unveiled the origin of the cosmic GeV gamma-ray background which has been a long standing problem in astrophysics. It is composed of blazars, radio galaxies, and star-forming galaxies. At $>100$~GeV, blazars dominate the background flux. It should be noted that estimation of contributions of radio galaxies and star-forming galaxies relies on limited samples. This situation is similar to blazar studies in the early EGRET era. Moreover, SEDs at the TeV gamma-ray band of these three populations are not fully understood. Future observational data will give deeper understanding on the cosmic GeV gamma-ray background.

Although {\it Fermi} has unveiled the origins of the cosmic GeV gamma-ray background, there are still unsolved questions. Those can be categorized to the following three theme; dark matter contribution, the cosmic MeV gamma-ray background, and the cosmic TeV gamma-ray background. These problems can be probed as follows. First, anisotropy of the cosmic GeV gamma-ray background will be useful to constrain the dark matter properties. Furthermore, cross correlation between the cosmic shear and the gamma-ray sky will be also a powerful probe of signatures of dark matter annihilation. Especially, cross correlation studies with coming optical wide field surveys will put tight constraints on dark matter properties. Second, anisotropy will also the key to understanding the origin of the cosmic MeV gamma-ray background, since it reflects the source distribution in the sky. Lastly, the cosmic TeV gamma-ray background has been already constrained by itself at the GeV gamma-ray band because the secondary gamma rays can not overproduce the measured gamma-ray background flux. More interestingly, if extragalactic $pp$ scenario is responsible for the recent IceCube neutrino events, they will contribute 30--100\% of the cosmic gamma-ray background at $\gtrsim100$~GeV, which is inconsistent with the expected blazars' contribution ($\sim$100\%)  at this energy band. Further detailed comparison between {\it Fermi} and IceCube would be important to understand the origin of the cosmic TeV gamma-ray and the TeV--PeV neutrino background.

\bigskip 
\begin{acknowledgments}
The author acknowledges support by the Research Fellowship of the Japan Society for the Promotion of Science and the JAXA international top young fellowship. The author wish to thank JACoW for their guidance in preparing this template. This work supported by Department of Energy contract DE-AC03-76SF00515.
\end{acknowledgments}

\bigskip 

\bibliography{Proceedings_LaTeX_YI}

\begin{thebibliography}{109}
\expandafter\ifx\csname natexlab\endcsname\relax\def\natexlab#1{#1}\fi
\expandafter\ifx\csname bibnamefont\endcsname\relax
  \def\bibnamefont#1{#1}\fi
\expandafter\ifx\csname bibfnamefont\endcsname\relax
  \def\bibfnamefont#1{#1}\fi
\expandafter\ifx\csname citenamefont\endcsname\relax
  \def\citenamefont#1{#1}\fi
\expandafter\ifx\csname url\endcsname\relax
  \def\url#1{\texttt{#1}}\fi
\expandafter\ifx\csname urlprefix\endcsname\relax\def\urlprefix{URL }\fi
\providecommand{\bibinfo}[2]{#2}
\providecommand{\eprint}[2][]{\url{#2}}

\bibitem{ued14}
\bibinfo{author}{\bibfnamefont{Y.}~\bibnamefont{{Ueda}}},
  \bibinfo{author}{\bibfnamefont{M.}~\bibnamefont{{Akiyama}}},
  \bibinfo{author}{\bibfnamefont{G.}~\bibnamefont{{Hasinger}}},
  \bibinfo{author}{\bibfnamefont{T.}~\bibnamefont{{Miyaji}}}, \bibnamefont{and}
  \bibinfo{author}{\bibfnamefont{M.~G.} \bibnamefont{{Watson}}},
  \bibinfo{journal}{\apj} \textbf{\bibinfo{volume}{786}}, \bibinfo{eid}{104}
  (\bibinfo{year}{2014}).

\bibitem{fic77} C.~E.~Fichtel et al.\ \apjl, 217, L9 (1977)

\bibitem{fic78}  C.~E.~Fichtel, G.~A.~Simpson, \& D.~J.~Thompson, \apj, 222, 833 (1978).



\bibitem{sre98}
\bibinfo{author}{\bibfnamefont{P.}~\bibnamefont{{Sreekumar}}}
  \bibnamefont{et~al.}, \bibinfo{journal}{\apj} \textbf{\bibinfo{volume}{494}},
  \bibinfo{pages}{523} (\bibinfo{year}{1998}).

\bibitem{str04}
\bibinfo{author}{\bibfnamefont{A.~W.} \bibnamefont{{Strong}}},
  \bibinfo{author}{\bibfnamefont{I.~V.} \bibnamefont{{Moskalenko}}},
  \bibnamefont{and} \bibinfo{author}{\bibfnamefont{O.}~\bibnamefont{{Reimer}}},
  \bibinfo{journal}{\apj} \textbf{\bibinfo{volume}{613}}, \bibinfo{pages}{956}
  (\bibinfo{year}{2004}).

\bibitem{ste08}
\bibinfo{author}{\bibfnamefont{F.~W.} \bibnamefont{{Stecker}}},
  \bibinfo{author}{\bibfnamefont{S.~D.} \bibnamefont{{Hunter}}},
  \bibnamefont{and} \bibinfo{author}{\bibfnamefont{D.~A.}
  \bibnamefont{{Kniffen}}}, \bibinfo{journal}{Astroparticle Physics}
  \textbf{\bibinfo{volume}{29}}, \bibinfo{pages}{25} (\bibinfo{year}{2008}).

\bibitem{ino09}
\bibinfo{author}{\bibfnamefont{Y.}~\bibnamefont{{Inoue}}} \bibnamefont{and}
  \bibinfo{author}{\bibfnamefont{T.}~\bibnamefont{{Totani}}},
  \bibinfo{journal}{\apj} \textbf{\bibinfo{volume}{702}}, \bibinfo{eid}{523}
  (\bibinfo{year}{2009}).

\bibitem{ino13_cib}
\bibinfo{author}{\bibfnamefont{Y.}~\bibnamefont{{Inoue}}},
  \bibinfo{author}{\bibfnamefont{S.}~\bibnamefont{{Inoue}}},
  \bibinfo{author}{\bibfnamefont{M.~A.~R.} \bibnamefont{{Kobayashi}}},
  \bibinfo{author}{\bibfnamefont{R.}~\bibnamefont{{Makiya}}},
  \bibinfo{author}{\bibfnamefont{Y.}~\bibnamefont{{Niino}}}, \bibnamefont{and}
  \bibinfo{author}{\bibfnamefont{T.}~\bibnamefont{{Totani}}},
  \bibinfo{journal}{\apj} \textbf{\bibinfo{volume}{768}}, \bibinfo{eid}{197}
  (\bibinfo{year}{2013}{\natexlab{a}}).

\bibitem{ino08}
\bibinfo{author}{\bibfnamefont{Y.}~\bibnamefont{{Inoue}}},
  \bibinfo{author}{\bibfnamefont{T.}~\bibnamefont{{Totani}}}, \bibnamefont{and}
  \bibinfo{author}{\bibfnamefont{Y.}~\bibnamefont{{Ueda}}},
  \bibinfo{journal}{\apjl} \textbf{\bibinfo{volume}{672}}, \bibinfo{pages}{L5}
  (\bibinfo{year}{2008}), \eprint{0709.3877}.

\bibitem{ino11}
\bibinfo{author}{\bibfnamefont{Y.}~\bibnamefont{{Inoue}}},
  \bibinfo{journal}{\apj} \textbf{\bibinfo{volume}{733}}, \bibinfo{eid}{66}
  (\bibinfo{year}{2011}), \eprint{1103.3946}.

\bibitem{ino13_cxb}
\bibinfo{author}{\bibfnamefont{Y.}~\bibnamefont{{Inoue}}},
  \bibinfo{author}{\bibfnamefont{K.}~\bibnamefont{{Murase}}},
  \bibinfo{author}{\bibfnamefont{G.~M.} \bibnamefont{{Madejski}}},
  \bibnamefont{and}
  \bibinfo{author}{\bibfnamefont{Y.}~\bibnamefont{{Uchiyama}}},
  \bibinfo{journal}{\apj} \textbf{\bibinfo{volume}{776}}, \bibinfo{eid}{33}
  (\bibinfo{year}{2013}{\natexlab{b}}).

\bibitem{ack14_cgb}
\bibinfo{author}{\bibfnamefont{M.}~\bibnamefont{{Ackermann}}}
  \bibnamefont{et~al.}, \bibinfo{journal}{arXiv:1410.3696}
  (\bibinfo{year}{2014}), \eprint{1410.3696}.

\bibitem{abd10_cgb}
\bibinfo{author}{\bibfnamefont{A.~A.} \bibnamefont{{Abdo}}}
  \bibnamefont{et~al.}, \bibinfo{journal}{Physical Review Letters}
  \textbf{\bibinfo{volume}{104}}, \bibinfo{eid}{101101}
  (\bibinfo{year}{2010}{\natexlab{a}}).

\bibitem{fun13}
\bibinfo{author}{\bibfnamefont{S.}~\bibnamefont{{Funk}}},
  \bibinfo{author}{\bibfnamefont{J.~A.} \bibnamefont{{Hinton}}},
  \bibinfo{author}{\bibnamefont{{for the CTA Consortium}}},
  \bibinfo{journal}{Astroparticle Physics} \textbf{\bibinfo{volume}{43}},
  \bibinfo{pages}{348} (\bibinfo{year}{2013}).

\bibitem{pad93}
\bibinfo{author}{\bibfnamefont{P.}~\bibnamefont{{Padovani}}},
  \bibinfo{author}{\bibfnamefont{G.}~\bibnamefont{{Ghisellini}}},
  \bibinfo{author}{\bibfnamefont{A.~C.} \bibnamefont{{Fabian}}},
  \bibnamefont{and}
  \bibinfo{author}{\bibfnamefont{A.}~\bibnamefont{{Celotti}}},
  \bibinfo{journal}{\mnras} \textbf{\bibinfo{volume}{260}},
  \bibinfo{pages}{L21} (\bibinfo{year}{1993}).

\bibitem{ste93}
\bibinfo{author}{\bibfnamefont{F.~W.} \bibnamefont{{Stecker}}},
  \bibinfo{author}{\bibfnamefont{M.~H.} \bibnamefont{{Salamon}}},
  \bibnamefont{and} \bibinfo{author}{\bibfnamefont{M.~A.}
  \bibnamefont{{Malkan}}}, \bibinfo{journal}{\apjl}
  \textbf{\bibinfo{volume}{410}}, \bibinfo{pages}{L71} (\bibinfo{year}{1993}).

\bibitem{str76}
\bibinfo{author}{\bibfnamefont{A.~W.} \bibnamefont{{Strong}}},
  \bibinfo{author}{\bibfnamefont{A.~W.} \bibnamefont{{Wolfendale}}},
  \bibnamefont{and} \bibinfo{author}{\bibfnamefont{D.~M.}
  \bibnamefont{{Worrall}}}, \bibinfo{journal}{\mnras}
  \textbf{\bibinfo{volume}{175}}, \bibinfo{pages}{23P} (\bibinfo{year}{1976}).

\bibitem{pav02}
\bibinfo{author}{\bibfnamefont{V.}~\bibnamefont{{Pavlidou}}} \bibnamefont{and}
  \bibinfo{author}{\bibfnamefont{B.~D.} \bibnamefont{{Fields}}},
  \bibinfo{journal}{\apjl} \textbf{\bibinfo{volume}{575}}, \bibinfo{pages}{L5}
  (\bibinfo{year}{2002}).

\bibitem{cas07}
\bibinfo{author}{\bibfnamefont{S.}~\bibnamefont{{Casanova}}},
  \bibinfo{author}{\bibfnamefont{B.~L.} \bibnamefont{{Dingus}}},
  \bibnamefont{and} \bibinfo{author}{\bibfnamefont{B.}~\bibnamefont{{Zhang}}},
  \bibinfo{journal}{\apj} \textbf{\bibinfo{volume}{656}}, \bibinfo{pages}{306}
  (\bibinfo{year}{2007}).

\bibitem{fau10}
\bibinfo{author}{\bibfnamefont{C.-A.} \bibnamefont{{Faucher-Gigu{\`e}re}}}
  \bibnamefont{and} \bibinfo{author}{\bibfnamefont{A.}~\bibnamefont{{Loeb}}},
  \bibinfo{journal}{\jcap} \textbf{\bibinfo{volume}{1}}, \bibinfo{eid}{005}
  (\bibinfo{year}{2010}).

\bibitem{loe00}
\bibinfo{author}{\bibfnamefont{A.}~\bibnamefont{{Loeb}}} \bibnamefont{and}
  \bibinfo{author}{\bibfnamefont{E.}~\bibnamefont{{Waxman}}},
  \bibinfo{journal}{\nat} \textbf{\bibinfo{volume}{405}}, \bibinfo{pages}{156}
  (\bibinfo{year}{2000}).

\bibitem{tot00}
\bibinfo{author}{\bibfnamefont{T.}~\bibnamefont{{Totani}}} \bibnamefont{and}
  \bibinfo{author}{\bibfnamefont{T.}~\bibnamefont{{Kitayama}}},
  \bibinfo{journal}{\apj} \textbf{\bibinfo{volume}{545}}, \bibinfo{pages}{572}
  (\bibinfo{year}{2000}).

\bibitem{dar95}
\bibinfo{author}{\bibfnamefont{A.}~\bibnamefont{{Dar}}} \bibnamefont{and}
  \bibinfo{author}{\bibfnamefont{N.~J.} \bibnamefont{{Shaviv}}},
  \bibinfo{journal}{Physical Review Letters} \textbf{\bibinfo{volume}{75}},
  \bibinfo{pages}{3052} (\bibinfo{year}{1995}).

\bibitem{kal09}
\bibinfo{author}{\bibfnamefont{O.~E.} \bibnamefont{{Kalashev}}},
  \bibinfo{author}{\bibfnamefont{D.~V.} \bibnamefont{{Semikoz}}},
  \bibnamefont{and} \bibinfo{author}{\bibfnamefont{G.}~\bibnamefont{{Sigl}}},
  \bibinfo{journal}{\prd} \textbf{\bibinfo{volume}{79}}, \bibinfo{eid}{063005}
  (\bibinfo{year}{2009}).

\bibitem{kes04}
\bibinfo{author}{\bibfnamefont{U.}~\bibnamefont{{Keshet}}},
  \bibinfo{author}{\bibfnamefont{E.}~\bibnamefont{{Waxman}}}, \bibnamefont{and}
  \bibinfo{author}{\bibfnamefont{A.}~\bibnamefont{{Loeb}}},
  \bibinfo{journal}{\jcap} \textbf{\bibinfo{volume}{4}}, \bibinfo{eid}{006}
  (\bibinfo{year}{2004}).

\bibitem{mos09}
\bibinfo{author}{\bibfnamefont{I.~V.} \bibnamefont{{Moskalenko}}}
  \bibnamefont{and} \bibinfo{author}{\bibfnamefont{T.~A.}
  \bibnamefont{{Porter}}}, \bibinfo{journal}{\apjl}
  \textbf{\bibinfo{volume}{692}}, \bibinfo{pages}{L54} (\bibinfo{year}{2009}).

\bibitem{ber01} Bergstr{\"o}m, L., Edsj{\"o}, J., \& Ullio, P., {\it PRL}  \textbf{87}, 251301 (2001).


\bibitem{sie11}
\bibinfo{author}{\bibfnamefont{J.~M.} \bibnamefont{{Siegal-Gaskins}}},
  \bibinfo{author}{\bibfnamefont{R.}~\bibnamefont{{Reesman}}},
  \bibinfo{author}{\bibfnamefont{V.}~\bibnamefont{{Pavlidou}}},
  \bibinfo{author}{\bibfnamefont{S.}~\bibnamefont{{Profumo}}},
  \bibnamefont{and} \bibinfo{author}{\bibfnamefont{T.~P.}
  \bibnamefont{{Walker}}}, \bibinfo{journal}{\mnras}
  \textbf{\bibinfo{volume}{415}}, \bibinfo{pages}{1074} (\bibinfo{year}{2011}).

\bibitem{aje14}
\bibinfo{author}{\bibfnamefont{M.}~\bibnamefont{{Ajello}}}
  \bibnamefont{et~al.}, \bibinfo{journal}{\apj} \textbf{\bibinfo{volume}{780}},
  \bibinfo{eid}{73} (\bibinfo{year}{2014}).

\bibitem{ree67}
\bibinfo{author}{\bibfnamefont{M.~J.} \bibnamefont{{Rees}}},
  \bibinfo{journal}{\mnras} \textbf{\bibinfo{volume}{137}},
  \bibinfo{pages}{429} (\bibinfo{year}{1967}).

\bibitem{mar92}
\bibinfo{author}{\bibfnamefont{L.}~\bibnamefont{{Maraschi}}},
  \bibinfo{author}{\bibfnamefont{G.}~\bibnamefont{{Ghisellini}}},
  \bibnamefont{and}
  \bibinfo{author}{\bibfnamefont{A.}~\bibnamefont{{Celotti}}},
  \bibinfo{journal}{\apjl} \textbf{\bibinfo{volume}{397}}, \bibinfo{pages}{L5}
  (\bibinfo{year}{1992}).

\bibitem{der93}
\bibinfo{author}{\bibfnamefont{C.~D.} \bibnamefont{{Dermer}}} \bibnamefont{and}
  \bibinfo{author}{\bibfnamefont{R.}~\bibnamefont{{Schlickeiser}}},
  \bibinfo{journal}{\apj} \textbf{\bibinfo{volume}{416}}, \bibinfo{pages}{458}
  (\bibinfo{year}{1993}).

\bibitem{sik94}
\bibinfo{author}{\bibfnamefont{M.}~\bibnamefont{{Sikora}}},
  \bibinfo{author}{\bibfnamefont{M.~C.} \bibnamefont{{Begelman}}},
  \bibnamefont{and} \bibinfo{author}{\bibfnamefont{M.~J.}
  \bibnamefont{{Rees}}}, \bibinfo{journal}{\apj}
  \textbf{\bibinfo{volume}{421}}, \bibinfo{pages}{153} (\bibinfo{year}{1994}).

\bibitem{man89}
\bibinfo{author}{\bibfnamefont{K.}~\bibnamefont{{Mannheim}}} \bibnamefont{and}
  \bibinfo{author}{\bibfnamefont{P.~L.} \bibnamefont{{Biermann}}},
  \bibinfo{journal}{\aap} \textbf{\bibinfo{volume}{221}}, \bibinfo{pages}{211}
  (\bibinfo{year}{1989}).

\bibitem{muc03}
\bibinfo{author}{\bibfnamefont{A.}~\bibnamefont{{M{\"u}cke}}},
  \bibinfo{author}{\bibfnamefont{R.~J.} \bibnamefont{{Protheroe}}},
  \bibinfo{author}{\bibfnamefont{R.}~\bibnamefont{{Engel}}},
  \bibinfo{author}{\bibfnamefont{J.~P.} \bibnamefont{{Rachen}}},
  \bibnamefont{and} \bibinfo{author}{\bibfnamefont{T.}~\bibnamefont{{Stanev}}},
  \bibinfo{journal}{Astroparticle Physics} \textbf{\bibinfo{volume}{18}},
  \bibinfo{pages}{593} (\bibinfo{year}{2003}).

\bibitem{fos98}
\bibinfo{author}{\bibfnamefont{G.}~\bibnamefont{{Fossati}}},
  \bibinfo{author}{\bibfnamefont{L.}~\bibnamefont{{Maraschi}}},
  \bibinfo{author}{\bibfnamefont{A.}~\bibnamefont{{Celotti}}},
  \bibinfo{author}{\bibfnamefont{A.}~\bibnamefont{{Comastri}}},
  \bibnamefont{and}
  \bibinfo{author}{\bibfnamefont{G.}~\bibnamefont{{Ghisellini}}},
  \bibinfo{journal}{\mnras} \textbf{\bibinfo{volume}{299}},
  \bibinfo{pages}{433} (\bibinfo{year}{1998}).

\bibitem{kub98}
\bibinfo{author}{\bibfnamefont{H.}~\bibnamefont{{Kubo}}},
  \bibinfo{author}{\bibfnamefont{T.}~\bibnamefont{{Takahashi}}},
  \bibinfo{author}{\bibfnamefont{G.}~\bibnamefont{{Madejski}}},
  \bibinfo{author}{\bibfnamefont{M.}~\bibnamefont{{Tashiro}}},
  \bibinfo{author}{\bibfnamefont{F.}~\bibnamefont{{Makino}}},
  \bibinfo{author}{\bibfnamefont{S.}~\bibnamefont{{Inoue}}}, \bibnamefont{and}
  \bibinfo{author}{\bibfnamefont{F.}~\bibnamefont{{Takahara}}},
  \bibinfo{journal}{\apj} \textbf{\bibinfo{volume}{504}}, \bibinfo{pages}{693}
  (\bibinfo{year}{1998}).

\bibitem{har99}
\bibinfo{author}{\bibfnamefont{R.~C.} \bibnamefont{{Hartman}}}
  \bibnamefont{et~al.}, \bibinfo{journal}{\apjs}
  \textbf{\bibinfo{volume}{123}}, \bibinfo{pages}{79} (\bibinfo{year}{1999}).

\bibitem{nol12}
\bibinfo{author}{\bibfnamefont{P.~L.} \bibnamefont{{Nolan}}}
  \bibnamefont{et~al.}, \bibinfo{journal}{\apjs}
  \textbf{\bibinfo{volume}{199}}, \bibinfo{eid}{31} (\bibinfo{year}{2012}).

\bibitem{aje12}
\bibinfo{author}{\bibfnamefont{M.}~\bibnamefont{{Ajello}}}
  \bibnamefont{et~al.}, \bibinfo{journal}{\apj} \textbf{\bibinfo{volume}{751}},
  \bibinfo{eid}{108} (\bibinfo{year}{2012}).

\bibitem{nar06}
\bibinfo{author}{\bibfnamefont{T.}~\bibnamefont{{Narumoto}}} \bibnamefont{and}
  \bibinfo{author}{\bibfnamefont{T.}~\bibnamefont{{Totani}}},
  \bibinfo{journal}{\apj} \textbf{\bibinfo{volume}{643}}, \bibinfo{pages}{81}
  (\bibinfo{year}{2006}).

\bibitem{hop06}
\bibinfo{author}{\bibfnamefont{A.~M.} \bibnamefont{{Hopkins}}}
  \bibnamefont{and} \bibinfo{author}{\bibfnamefont{J.~F.}
  \bibnamefont{{Beacom}}}, \bibinfo{journal}{\apj}
  \textbf{\bibinfo{volume}{651}}, \bibinfo{pages}{142} (\bibinfo{year}{2006}).

\bibitem{gio99}
\bibinfo{author}{\bibfnamefont{P.}~\bibnamefont{{Giommi}}},
  \bibinfo{author}{\bibfnamefont{M.~T.} \bibnamefont{{Menna}}},
  \bibnamefont{and}
  \bibinfo{author}{\bibfnamefont{P.}~\bibnamefont{{Padovani}}},
  \bibinfo{journal}{\mnras} \textbf{\bibinfo{volume}{310}},
  \bibinfo{pages}{465} (\bibinfo{year}{1999}).

\bibitem{pad07}
\bibinfo{author}{\bibfnamefont{P.}~\bibnamefont{{Padovani}}},
  \bibinfo{author}{\bibfnamefont{P.}~\bibnamefont{{Giommi}}},
  \bibinfo{author}{\bibfnamefont{H.}~\bibnamefont{{Landt}}}, \bibnamefont{and}
  \bibinfo{author}{\bibfnamefont{E.~S.} \bibnamefont{{Perlman}}},
  \bibinfo{journal}{\apj} \textbf{\bibinfo{volume}{662}}, \bibinfo{pages}{182}
  (\bibinfo{year}{2007}).

\bibitem{mar13}
\bibinfo{author}{\bibfnamefont{M.~J.~M.} \bibnamefont{{March{\~a}}}}
  \bibnamefont{and}
  \bibinfo{author}{\bibfnamefont{A.}~\bibnamefont{{Caccianiga}}},
  \bibinfo{journal}{\mnras} \textbf{\bibinfo{volume}{430}},
  \bibinfo{pages}{2464} (\bibinfo{year}{2013}).

\bibitem{abd10_magn}
\bibinfo{author}{\bibfnamefont{A.~A.} \bibnamefont{{Abdo}}}
  \bibnamefont{et~al.}, \bibinfo{journal}{\apj} \textbf{\bibinfo{volume}{720}},
  \bibinfo{pages}{912} (\bibinfo{year}{2010}{\natexlab{b}}).

\bibitem{wil01}
\bibinfo{author}{\bibfnamefont{C.~J.} \bibnamefont{{Willott}}},
  \bibinfo{author}{\bibfnamefont{S.}~\bibnamefont{{Rawlings}}},
  \bibinfo{author}{\bibfnamefont{K.~M.} \bibnamefont{{Blundell}}},
  \bibinfo{author}{\bibfnamefont{M.}~\bibnamefont{{Lacy}}}, \bibnamefont{and}
  \bibinfo{author}{\bibfnamefont{S.~A.} \bibnamefont{{Eales}}},
  \bibinfo{journal}{\mnras} \textbf{\bibinfo{volume}{322}},
  \bibinfo{pages}{536} (\bibinfo{year}{2001}).

\bibitem{dim14}
\bibinfo{author}{\bibfnamefont{M.}~\bibnamefont{{Di Mauro}}},
  \bibinfo{author}{\bibfnamefont{F.}~\bibnamefont{{Calore}}},
  \bibinfo{author}{\bibfnamefont{F.}~\bibnamefont{{Donato}}},
  \bibinfo{author}{\bibfnamefont{M.}~\bibnamefont{{Ajello}}}, \bibnamefont{and}
  \bibinfo{author}{\bibfnamefont{L.}~\bibnamefont{{Latronico}}},
  \bibinfo{journal}{\apj} \textbf{\bibinfo{volume}{780}}, \bibinfo{eid}{161}
  (\bibinfo{year}{2014}{\natexlab{a}}), \eprint{1304.0908}.

\bibitem{ack12}
\bibinfo{author}{\bibfnamefont{M.}~\bibnamefont{{Ackermann}}}
  \bibnamefont{et~al.}, \bibinfo{journal}{\apj} \textbf{\bibinfo{volume}{755}},
  \bibinfo{eid}{164} (\bibinfo{year}{2012}{\natexlab{a}}).

\bibitem{ste96}
\bibinfo{author}{\bibfnamefont{F.~W.} \bibnamefont{{Stecker}}}
  \bibnamefont{and} \bibinfo{author}{\bibfnamefont{M.~H.}
  \bibnamefont{{Salamon}}}, \bibinfo{journal}{\apj}
  \textbf{\bibinfo{volume}{464}}, \bibinfo{pages}{600} (\bibinfo{year}{1996}).

\bibitem{aje14_chicago}
\bibinfo{author}{\bibfnamefont{M.}~\bibnamefont{{Ajello}}}
  \bibnamefont{et~al.}, \bibinfo{journal}{High Energy Messengers: Connecting the Non-Thermal Extragalactic Backgrounds}
  (\bibinfo{year}{2014}) {\url{https://kicp-workshops.uchicago.edu/hem2014/}}.

\bibitem{ack12_anis}
\bibinfo{author}{\bibfnamefont{M.}~\bibnamefont{{Ackermann}}}
  \bibnamefont{et~al.}, \bibinfo{journal}{\prd} \textbf{\bibinfo{volume}{85}},
  \bibinfo{eid}{083007} (\bibinfo{year}{2012}{\natexlab{b}}).

\bibitem{and06}
\bibinfo{author}{\bibfnamefont{S.}~\bibnamefont{{Ando}}} \bibnamefont{and}
  \bibinfo{author}{\bibfnamefont{E.}~\bibnamefont{{Komatsu}}},
  \bibinfo{journal}{\prd} \textbf{\bibinfo{volume}{73}}, \bibinfo{eid}{023521}
  (\bibinfo{year}{2006}).

\bibitem{and07}
\bibinfo{author}{\bibfnamefont{S.}~\bibnamefont{{Ando}}},
  \bibinfo{author}{\bibfnamefont{E.}~\bibnamefont{{Komatsu}}},
  \bibinfo{author}{\bibfnamefont{T.}~\bibnamefont{{Narumoto}}},
  \bibnamefont{and} \bibinfo{author}{\bibfnamefont{T.}~\bibnamefont{{Totani}}},
  \bibinfo{journal}{\prd} \textbf{\bibinfo{volume}{75}}, \bibinfo{eid}{063519}
  (\bibinfo{year}{2007}).

\bibitem{and13}
\bibinfo{author}{\bibfnamefont{S.}~\bibnamefont{{Ando}}} \bibnamefont{and}
  \bibinfo{author}{\bibfnamefont{E.}~\bibnamefont{{Komatsu}}},
  \bibinfo{journal}{\prd} \textbf{\bibinfo{volume}{87}}, \bibinfo{eid}{123539}
  (\bibinfo{year}{2013}).

\bibitem{shi14}
\bibinfo{author}{\bibfnamefont{M.}~\bibnamefont{{Shirasaki}}},
  \bibinfo{author}{\bibfnamefont{S.}~\bibnamefont{{Horiuchi}}},
  \bibnamefont{and}
  \bibinfo{author}{\bibfnamefont{N.}~\bibnamefont{{Yoshida}}},
  \bibinfo{journal}{\prd} \textbf{\bibinfo{volume}{90}}, \bibinfo{eid}{063502}
  (\bibinfo{year}{2014}).


\bibitem{cam13}
\bibinfo{author}{\bibfnamefont{S.}~\bibnamefont{{Camera}}},
  \bibinfo{author}{\bibfnamefont{M.}~\bibnamefont{{Fornasa}}},
  \bibinfo{author}{\bibfnamefont{N.}~\bibnamefont{{Fornengo}}},
  \bibnamefont{and} \bibinfo{author}{\bibfnamefont{M.}~\bibnamefont{{Regis}}},
  \bibinfo{journal}{\apjl} \textbf{\bibinfo{volume}{771}}, \bibinfo{eid}{L5}
  (\bibinfo{year}{2013}).


\bibitem{cuo12}
\bibinfo{author}{\bibfnamefont{A.}~\bibnamefont{{Cuoco}}},
  \bibinfo{author}{\bibfnamefont{E.}~\bibnamefont{{Komatsu}}},
  \bibnamefont{and} \bibinfo{author}{\bibfnamefont{J.~M.}
  \bibnamefont{{Siegal-Gaskins}}}, \bibinfo{journal}{\prd}
  \textbf{\bibinfo{volume}{86}}, \bibinfo{eid}{063004} (\bibinfo{year}{2012}).

\bibitem{har12}
\bibinfo{author}{\bibfnamefont{J.~P.} \bibnamefont{{Harding}}}
  \bibnamefont{and} \bibinfo{author}{\bibfnamefont{K.~N.}
  \bibnamefont{{Abazajian}}}, \bibinfo{journal}{\jcap}
  \textbf{\bibinfo{volume}{11}}, \bibinfo{eid}{026} (\bibinfo{year}{2012}).

\bibitem{dim14_anis}
\bibinfo{author}{\bibfnamefont{M.}~\bibnamefont{{Di Mauro}}},
  \bibinfo{author}{\bibfnamefont{A.}~\bibnamefont{{Cuoco}}},
  \bibinfo{author}{\bibfnamefont{F.}~\bibnamefont{{Donato}}}, \bibnamefont{and}
  \bibinfo{author}{\bibfnamefont{J.~M.} \bibnamefont{{Siegal-Gaskins}}},
  \bibinfo{journal}{ArXiv e-prints}  (\bibinfo{year}{2014}{\natexlab{b}}),
  \eprint{1407.3275}.

\bibitem{fuk75}
\bibinfo{author}{\bibfnamefont{Y.}~\bibnamefont{{Fukada}}}
  \bibnamefont{et~al.}, \bibinfo{journal}{\nat} \textbf{\bibinfo{volume}{254}},
  \bibinfo{pages}{398} (\bibinfo{year}{1975}).

\bibitem{wat97}
\bibinfo{author}{\bibfnamefont{K.}~\bibnamefont{{Watanabe}}}
  \bibnamefont{et~al.}, in \emph{\bibinfo{booktitle}{Proceedings of the Fourth
  Compton Symposium}}, edited by \bibinfo{editor}{\bibfnamefont{C.~D.}
  \bibnamefont{{Dermer}}}, \bibinfo{editor}{\bibfnamefont{M.~S.}
  \bibnamefont{{Strickman}}}, \bibnamefont{and}
  \bibinfo{editor}{\bibfnamefont{J.~D.} \bibnamefont{{Kurfess}}}
  (\bibinfo{year}{1997}), vol. \bibinfo{volume}{410} of
  \emph{\bibinfo{series}{American Institute of Physics Conference Series}}, pp.
  \bibinfo{pages}{1223--1227}.

\bibitem{wei00}
\bibinfo{author}{\bibfnamefont{G.}~\bibnamefont{{Weidenspointner}}}
  \bibnamefont{et~al.}, in \emph{\bibinfo{booktitle}{American Institute of
  Physics Conference Series}}, edited by \bibinfo{editor}{\bibfnamefont{M.~L.}
  \bibnamefont{{McConnell}}} \bibnamefont{and}
  \bibinfo{editor}{\bibfnamefont{J.~M.} \bibnamefont{{Ryan}}}
  (\bibinfo{year}{2000}), vol. \bibinfo{volume}{510} of
  \emph{\bibinfo{series}{American Institute of Physics Conference Series}}, pp.
  \bibinfo{pages}{467--470}.

\bibitem{cla75}
\bibinfo{author}{\bibfnamefont{D.~D.} \bibnamefont{{Clayton}}}
  \bibnamefont{and} \bibinfo{author}{\bibfnamefont{R.~A.}
  \bibnamefont{{Ward}}}, \bibinfo{journal}{\apj}
  \textbf{\bibinfo{volume}{198}}, \bibinfo{pages}{241} (\bibinfo{year}{1975}).

\bibitem{ahn05_sn}
\bibinfo{author}{\bibfnamefont{K.}~\bibnamefont{{Ahn}}},
  \bibinfo{author}{\bibfnamefont{E.}~\bibnamefont{{Komatsu}}},
  \bibnamefont{and}
  \bibinfo{author}{\bibfnamefont{P.}~\bibnamefont{{H{\"o}flich}}},
  \bibinfo{journal}{\prd} \textbf{\bibinfo{volume}{71}}, \bibinfo{eid}{121301}
  (\bibinfo{year}{2005}).

\bibitem{hor10}
\bibinfo{author}{\bibfnamefont{S.}~\bibnamefont{{Horiuchi}}} \bibnamefont{and}
  \bibinfo{author}{\bibfnamefont{J.~F.} \bibnamefont{{Beacom}}},
  \bibinfo{journal}{\apj} \textbf{\bibinfo{volume}{723}}, \bibinfo{pages}{329}
  (\bibinfo{year}{2010}).

\bibitem{sch78}
\bibinfo{author}{\bibfnamefont{V.}~\bibnamefont{{Schoenfelder}}},
  \bibinfo{journal}{\nat} \textbf{\bibinfo{volume}{274}}, \bibinfo{pages}{344}
  (\bibinfo{year}{1978}).

\bibitem{aje09}
\bibinfo{author}{\bibfnamefont{M.}~\bibnamefont{{Ajello}}}
  \bibnamefont{et~al.}, \bibinfo{journal}{\apj} \textbf{\bibinfo{volume}{699}},
  \bibinfo{pages}{603} (\bibinfo{year}{2009}).

\bibitem{mas11}
\bibinfo{author}{\bibfnamefont{F.}~\bibnamefont{{Massaro}}} \bibnamefont{and}
  \bibinfo{author}{\bibfnamefont{M.}~\bibnamefont{{Ajello}}},
  \bibinfo{journal}{\apjl} \textbf{\bibinfo{volume}{729}}, \bibinfo{eid}{L12}
  (\bibinfo{year}{2011}).

\bibitem{oli85}
\bibinfo{author}{\bibfnamefont{K.~A.} \bibnamefont{{Olive}}} \bibnamefont{and}
  \bibinfo{author}{\bibfnamefont{J.}~\bibnamefont{{Silk}}},
  \bibinfo{journal}{Physical Review Letters} \textbf{\bibinfo{volume}{55}},
  \bibinfo{pages}{2362} (\bibinfo{year}{1985}).

\bibitem{ahn05_dm1}
\bibinfo{author}{\bibfnamefont{K.}~\bibnamefont{{Ahn}}} \bibnamefont{and}
  \bibinfo{author}{\bibfnamefont{E.}~\bibnamefont{{Komatsu}}},
  \bibinfo{journal}{\prd} \textbf{\bibinfo{volume}{71}}, \bibinfo{eid}{021303}
  (\bibinfo{year}{2005}{\natexlab{a}}).

\bibitem{ahn05_dm2}
\bibinfo{author}{\bibfnamefont{K.}~\bibnamefont{{Ahn}}} \bibnamefont{and}
  \bibinfo{author}{\bibfnamefont{E.}~\bibnamefont{{Komatsu}}},
  \bibinfo{journal}{\prd} \textbf{\bibinfo{volume}{72}}, \bibinfo{eid}{061301}
  (\bibinfo{year}{2005}{\natexlab{b}}).

\bibitem{ued03}
\bibinfo{author}{\bibfnamefont{Y.}~\bibnamefont{{Ueda}}},
  \bibinfo{author}{\bibfnamefont{M.}~\bibnamefont{{Akiyama}}},
  \bibinfo{author}{\bibfnamefont{K.}~\bibnamefont{{Ohta}}}, \bibnamefont{and}
  \bibinfo{author}{\bibfnamefont{T.}~\bibnamefont{{Miyaji}}},
  \bibinfo{journal}{\apj} \textbf{\bibinfo{volume}{598}}, \bibinfo{pages}{886}
  (\bibinfo{year}{2003}).

\bibitem{tak12}
\bibinfo{author}{\bibfnamefont{T.}~\bibnamefont{{Takahashi}}}
  \bibnamefont{et~al.}, in \emph{\bibinfo{booktitle}{Society of Photo-Optical
  Instrumentation Engineers (SPIE) Conference Series}} (\bibinfo{year}{2012}),
  vol. \bibinfo{volume}{8443} of \emph{\bibinfo{series}{Society of
  Photo-Optical Instrumentation Engineers (SPIE) Conference Series}},
  p.~\bibinfo{pages}{1}.

\bibitem{ino14_corona}
\bibinfo{author}{\bibfnamefont{Y.}~\bibnamefont{{Inoue}}} \bibnamefont{and}
  \bibinfo{author}{\bibfnamefont{A.}~\bibnamefont{{Doi}}},
  \bibinfo{journal}{\pasj}  (\bibinfo{year}{2014}), \eprint{1411.2334}.

\bibitem{ino12}
\bibinfo{author}{\bibfnamefont{Y.}~\bibnamefont{{Inoue}}} \bibnamefont{and}
  \bibinfo{author}{\bibfnamefont{K.}~\bibnamefont{{Ioka}}},
  \bibinfo{journal}{\prd} \textbf{\bibinfo{volume}{86}}, \bibinfo{eid}{023003}
  (\bibinfo{year}{2012}).

\bibitem{ack11_TeVPA}
\bibinfo{author}{\bibfnamefont{M.}~\bibnamefont{{Ackermann}}}
  \bibnamefont{et~al.}, \bibinfo{journal}{TeV Particle Astrophysics 2011}
  (\bibinfo{year}{2011}) {\url{http://agenda.albanova.se/conferenceDisplay.py?confId=2600}}.

\bibitem{cha14}
\bibinfo{author}{\bibfnamefont{P.}~\bibnamefont{{Chang}}},
  \bibinfo{author}{\bibfnamefont{A.~E.} \bibnamefont{{Broderick}}},
  \bibinfo{author}{\bibfnamefont{C.}~\bibnamefont{{Pfrommer}}},
  \bibinfo{author}{\bibfnamefont{E.}~\bibnamefont{{Puchwein}}},
  \bibinfo{author}{\bibfnamefont{A.}~\bibnamefont{{Lamberts}}},
  \bibnamefont{and}
  \bibinfo{author}{\bibfnamefont{M.}~\bibnamefont{{Shalaby}}},
  \bibinfo{journal}{ArXiv e-prints}  (\bibinfo{year}{2014}),
  \eprint{1410.3797}.

\bibitem{bro12}
\bibinfo{author}{\bibfnamefont{A.~E.} \bibnamefont{{Broderick}}},
  \bibinfo{author}{\bibfnamefont{P.}~\bibnamefont{{Chang}}}, \bibnamefont{and}
  \bibinfo{author}{\bibfnamefont{C.}~\bibnamefont{{Pfrommer}}},
  \bibinfo{journal}{\apj} \textbf{\bibinfo{volume}{752}}, \bibinfo{eid}{22}
  (\bibinfo{year}{2012}).

\bibitem{sir14}
\bibinfo{author}{\bibfnamefont{L.}~\bibnamefont{{Sironi}}} \bibnamefont{and}
  \bibinfo{author}{\bibfnamefont{D.}~\bibnamefont{{Giannios}}},
  \bibinfo{journal}{\apj} \textbf{\bibinfo{volume}{787}}, \bibinfo{eid}{49}
  (\bibinfo{year}{2014}).

\bibitem{cop97}
\bibinfo{author}{\bibfnamefont{P.~S.} \bibnamefont{{Coppi}}} \bibnamefont{and}
  \bibinfo{author}{\bibfnamefont{F.~A.} \bibnamefont{{Aharonian}}},
  \bibinfo{journal}{\apjl} \textbf{\bibinfo{volume}{487}}, \bibinfo{pages}{L9}
  (\bibinfo{year}{1997}), \eprint{astro-ph/9610176}.

\bibitem{mur12}
\bibinfo{author}{\bibfnamefont{K.}~\bibnamefont{{Murase}}},
  \bibinfo{author}{\bibfnamefont{J.~F.} \bibnamefont{{Beacom}}},
  \bibnamefont{and} \bibinfo{author}{\bibfnamefont{H.}~\bibnamefont{{Takami}}},
  \bibinfo{journal}{\jcap} \textbf{\bibinfo{volume}{8}}, \bibinfo{eid}{030}
  (\bibinfo{year}{2012}).

\bibitem{aar13}
\bibinfo{author}{\bibfnamefont{M.~G.} \bibnamefont{{Aartsen}}}
  \bibnamefont{et~al.}, \bibinfo{journal}{Physical Review Letters}
  \textbf{\bibinfo{volume}{111}}, \bibinfo{eid}{021103} (\bibinfo{year}{2013}).

\bibitem{mur13}
\bibinfo{author}{\bibfnamefont{K.}~\bibnamefont{{Murase}}},
  \bibinfo{author}{\bibfnamefont{M.}~\bibnamefont{{Ahlers}}}, \bibnamefont{and}
  \bibinfo{author}{\bibfnamefont{B.~C.} \bibnamefont{{Lacki}}},
  \bibinfo{journal}{\prd} \textbf{\bibinfo{volume}{88}}, \bibinfo{eid}{121301}
  (\bibinfo{year}{2013}).

\bibitem{aar14}
\bibinfo{author}{\bibfnamefont{M.~G.} \bibnamefont{{Aartsen}}}
  \bibnamefont{et~al.}, \bibinfo{journal}{Physical Review Letters}
  \textbf{\bibinfo{volume}{113}}, \bibinfo{eid}{101101} (\bibinfo{year}{2014}).

\bibitem{mur14}
\bibinfo{author}{\bibfnamefont{K.}~\bibnamefont{{Murase}}},
  \bibinfo{journal}{ArXiv e-prints}  (\bibinfo{year}{2014}),
  \eprint{1410.3680}.

\bibitem{mur14_seq}
\bibinfo{author}{\bibfnamefont{K.}~\bibnamefont{{Murase}}},
  \bibinfo{author}{\bibfnamefont{Y.}~\bibnamefont{{Inoue}}}, \bibnamefont{and}
  \bibinfo{author}{\bibfnamefont{C.~D.} \bibnamefont{{Dermer}}},
  \bibinfo{journal}{\prd} \textbf{\bibinfo{volume}{90}}, \bibinfo{eid}{023007}
  (\bibinfo{year}{2014}).

\bibitem{der14}
\bibinfo{author}{\bibfnamefont{C.~D.} \bibnamefont{{Dermer}}},
  \bibinfo{author}{\bibfnamefont{K.}~\bibnamefont{{Murase}}}, \bibnamefont{and}
  \bibinfo{author}{\bibfnamefont{Y.}~\bibnamefont{{Inoue}}},
  \bibinfo{journal}{Journal of High Energy Astrophysics}
  \textbf{\bibinfo{volume}{3}}, \bibinfo{pages}{29} (\bibinfo{year}{2014}).

\bibitem{dwe13}
\bibinfo{author}{\bibfnamefont{E.}~\bibnamefont{{Dwek}}} \bibnamefont{and}
  \bibinfo{author}{\bibfnamefont{F.}~\bibnamefont{{Krennrich}}},
  \bibinfo{journal}{Astroparticle Physics} \textbf{\bibinfo{volume}{43}},
  \bibinfo{pages}{112} (\bibinfo{year}{2013}).

\bibitem{dom11}
\bibinfo{author}{\bibfnamefont{A.}~\bibnamefont{{Dom{\'{\i}}nguez}}}
  \bibnamefont{et~al.}, \bibinfo{journal}{\mnras}
  \textbf{\bibinfo{volume}{410}}, \bibinfo{pages}{2556} (\bibinfo{year}{2011}).

\bibitem{mad00}
\bibinfo{author}{\bibfnamefont{P.}~\bibnamefont{{Madau}}} \bibnamefont{and}
  \bibinfo{author}{\bibfnamefont{L.}~\bibnamefont{{Pozzetti}}},
  \bibinfo{journal}{\mnras} \textbf{\bibinfo{volume}{312}}, \bibinfo{pages}{L9}
  (\bibinfo{year}{2000}).

\bibitem{tot01}
\bibinfo{author}{\bibfnamefont{T.}~\bibnamefont{{Totani}}},
  \bibinfo{author}{\bibfnamefont{Y.}~\bibnamefont{{Yoshii}}},
  \bibinfo{author}{\bibfnamefont{F.}~\bibnamefont{{Iwamuro}}},
  \bibinfo{author}{\bibfnamefont{T.}~\bibnamefont{{Maihara}}},
  \bibnamefont{and}
  \bibinfo{author}{\bibfnamefont{K.}~\bibnamefont{{Motohara}}},
  \bibinfo{journal}{\apjl} \textbf{\bibinfo{volume}{550}},
  \bibinfo{pages}{L137} (\bibinfo{year}{2001}).

\bibitem{mat05}
\bibinfo{author}{\bibfnamefont{T.}~\bibnamefont{{Matsumoto}}}
  \bibnamefont{et~al.}, \bibinfo{journal}{\apj} \textbf{\bibinfo{volume}{626}},
  \bibinfo{pages}{31} (\bibinfo{year}{2005}).

\bibitem{tsu13}
\bibinfo{author}{\bibfnamefont{K.}~\bibnamefont{{Tsumura}}},
  \bibinfo{author}{\bibfnamefont{T.}~\bibnamefont{{Matsumoto}}},
  \bibinfo{author}{\bibfnamefont{S.}~\bibnamefont{{Matsuura}}},
  \bibinfo{author}{\bibfnamefont{I.}~\bibnamefont{{Sakon}}}, \bibnamefont{and}
  \bibinfo{author}{\bibfnamefont{T.}~\bibnamefont{{Wada}}},
  \bibinfo{journal}{\pasj} \textbf{\bibinfo{volume}{65}}, \bibinfo{pages}{121}
  (\bibinfo{year}{2013}).

\bibitem{mad05}
\bibinfo{author}{\bibfnamefont{P.}~\bibnamefont{{Madau}}} \bibnamefont{and}
  \bibinfo{author}{\bibfnamefont{J.}~\bibnamefont{{Silk}}},
  \bibinfo{journal}{\mnras} \textbf{\bibinfo{volume}{359}},
  \bibinfo{pages}{L37} (\bibinfo{year}{2005}).

\bibitem{ino14}
\bibinfo{author}{\bibfnamefont{Y.}~\bibnamefont{{Inoue}}},
  \bibinfo{author}{\bibfnamefont{Y.~T.} \bibnamefont{{Tanaka}}},
  \bibinfo{author}{\bibfnamefont{G.~M.} \bibnamefont{{Madejski}}},
  \bibnamefont{and}
  \bibinfo{author}{\bibfnamefont{A.}~\bibnamefont{{Dom{\'{\i}}nguez}}},
  \bibinfo{journal}{\apjl} \textbf{\bibinfo{volume}{781}}, \bibinfo{eid}{L35}
  (\bibinfo{year}{2014}), \eprint{1312.6462}.

\bibitem{aha06}
\bibinfo{author}{\bibfnamefont{F.}~\bibnamefont{{Aharonian}}}
  \bibnamefont{et~al.}, \bibinfo{journal}{\nat} \textbf{\bibinfo{volume}{440}},
  \bibinfo{pages}{1018} (\bibinfo{year}{2006}).

\bibitem{ess10}
\bibinfo{author}{\bibfnamefont{W.}~\bibnamefont{{Essey}}} \bibnamefont{and}
  \bibinfo{author}{\bibfnamefont{A.}~\bibnamefont{{Kusenko}}},
  \bibinfo{journal}{Astroparticle Physics} \textbf{\bibinfo{volume}{33}},
  \bibinfo{pages}{81} (\bibinfo{year}{2010}).

\bibitem{mat11}
\bibinfo{author}{\bibfnamefont{Y.}~\bibnamefont{{Matsuoka}}},
  \bibinfo{author}{\bibfnamefont{N.}~\bibnamefont{{Ienaka}}},
  \bibinfo{author}{\bibfnamefont{K.}~\bibnamefont{{Kawara}}}, \bibnamefont{and}
  \bibinfo{author}{\bibfnamefont{S.}~\bibnamefont{{Oyabu}}},
  \bibinfo{journal}{\apj} \textbf{\bibinfo{volume}{736}}, \bibinfo{eid}{119}
  (\bibinfo{year}{2011}).

\bibitem{gou66}
\bibinfo{author}{\bibfnamefont{R.~J.} \bibnamefont{{Gould}}} \bibnamefont{and}
  \bibinfo{author}{\bibfnamefont{G.}~\bibnamefont{{Schr{\'e}der}}},
  \bibinfo{journal}{Physical Review Letters} \textbf{\bibinfo{volume}{16}},
  \bibinfo{pages}{252} (\bibinfo{year}{1966}).

\bibitem{jel66}
\bibinfo{author}{\bibfnamefont{J.~V.} \bibnamefont{{Jelley}}},
  \bibinfo{journal}{Physical Review Letters} \textbf{\bibinfo{volume}{16}},
  \bibinfo{pages}{479} (\bibinfo{year}{1966}).

\bibitem{ste92}
\bibinfo{author}{\bibfnamefont{F.~W.} \bibnamefont{{Stecker}}},
  \bibinfo{author}{\bibfnamefont{O.~C.} \bibnamefont{{de Jager}}},
  \bibnamefont{and} \bibinfo{author}{\bibfnamefont{M.~H.}
  \bibnamefont{{Salamon}}}, \bibinfo{journal}{\apjl}
  \textbf{\bibinfo{volume}{390}}, \bibinfo{pages}{L49} (\bibinfo{year}{1992}).

\bibitem{abd09_080916C}
\bibinfo{author}{\bibfnamefont{A.~A.} \bibnamefont{{Abdo}}}
  \bibnamefont{et~al.}, \bibinfo{journal}{Science}
  \textbf{\bibinfo{volume}{323}}, \bibinfo{pages}{1688} (\bibinfo{year}{2009}).

\bibitem[{\citenamefont{{Meyer} et~al.}(2012)\citenamefont{{Meyer}, {Raue},
  {Mazin}, and {Horns}}}]{mey12}
\bibinfo{author}{\bibfnamefont{M.}~\bibnamefont{{Meyer}}},
  \bibinfo{author}{\bibfnamefont{M.}~\bibnamefont{{Raue}}},
  \bibinfo{author}{\bibfnamefont{D.}~\bibnamefont{{Mazin}}}, \bibnamefont{and}
  \bibinfo{author}{\bibfnamefont{D.}~\bibnamefont{{Horns}}},
  \bibinfo{journal}{\aap} \textbf{\bibinfo{volume}{542}}, \bibinfo{eid}{A59}
  (\bibinfo{year}{2012}).

\bibitem{hor09}
\bibinfo{author}{\bibfnamefont{S.}~\bibnamefont{{Horiuchi}}},
  \bibinfo{author}{\bibfnamefont{J.~F.} \bibnamefont{{Beacom}}},
  \bibnamefont{and} \bibinfo{author}{\bibfnamefont{E.}~\bibnamefont{{Dwek}}},
  \bibinfo{journal}{\prd} \textbf{\bibinfo{volume}{79}}, \bibinfo{eid}{083013}
  (\bibinfo{year}{2009}).

\bibitem{abr13}
\bibinfo{author}{\bibfnamefont{A.}~\bibnamefont{{Abramowski}}}
  \bibnamefont{et~al.}, \bibinfo{journal}{\aap} \textbf{\bibinfo{volume}{550}},
  \bibinfo{eid}{A4} (\bibinfo{year}{2013}).

\bibitem{ack12_ebl}
\bibinfo{author}{\bibfnamefont{M.}~\bibnamefont{{Ackermann}}}
  \bibnamefont{et~al.}, \bibinfo{journal}{Science}
  \textbf{\bibinfo{volume}{338}}, \bibinfo{pages}{1190}
  (\bibinfo{year}{2012}{\natexlab{c}}).

\bibitem{lef11}
\bibinfo{author}{\bibfnamefont{E.}~\bibnamefont{{Lefa}}},
  \bibinfo{author}{\bibfnamefont{F.~M.} \bibnamefont{{Rieger}}},
  \bibnamefont{and}
  \bibinfo{author}{\bibfnamefont{F.}~\bibnamefont{{Aharonian}}},
  \bibinfo{journal}{\apj} \textbf{\bibinfo{volume}{740}}, \bibinfo{eid}{64}
  (\bibinfo{year}{2011}).

\bibitem{zem14}
\bibinfo{author}{\bibfnamefont{M.}~\bibnamefont{{Zemcov}}}
  \bibnamefont{et~al.}, \bibinfo{journal}{Science}
  \textbf{\bibinfo{volume}{346}}, \bibinfo{pages}{732} (\bibinfo{year}{2014}).

\bibitem{kas05}
\bibinfo{author}{\bibfnamefont{A.}~\bibnamefont{{Kashlinsky}}},
  \bibinfo{author}{\bibfnamefont{R.~G.} \bibnamefont{{Arendt}}},
  \bibinfo{author}{\bibfnamefont{J.}~\bibnamefont{{Mather}}}, \bibnamefont{and}
  \bibinfo{author}{\bibfnamefont{S.~H.} \bibnamefont{{Moseley}}},
  \bibinfo{journal}{\nat} \textbf{\bibinfo{volume}{438}}, \bibinfo{pages}{45}
  (\bibinfo{year}{2005}).

\bibitem{coo07}
\bibinfo{author}{\bibfnamefont{A.}~\bibnamefont{{Cooray}}}
  \bibnamefont{et~al.}, \bibinfo{journal}{\apjl}
  \textbf{\bibinfo{volume}{659}}, \bibinfo{pages}{L91} (\bibinfo{year}{2007}).

\bibitem{mat11_flu}
\bibinfo{author}{\bibfnamefont{T.}~\bibnamefont{{Matsumoto}}}
  \bibnamefont{et~al.}, \bibinfo{journal}{\apj} \textbf{\bibinfo{volume}{742}},
  \bibinfo{eid}{124} (\bibinfo{year}{2011}).

\end{thebibliography}





\end{document}